\documentclass[pra,aps,twocolumn,superscriptaddress,floatfix,showpacs,longbibliography]{revtex4-2}
\usepackage{graphicx}
\usepackage{dcolumn}
\usepackage{bm}
\usepackage{upgreek}
\usepackage{natbib}
\usepackage[normalem]{ulem}
\usepackage{amsmath,amssymb}
\usepackage{dsfont}
\usepackage[english]{babel}
\usepackage{braket}
\usepackage{color}
\usepackage{appendix}
\begin{document}
%\title{Ground State Phases and Topological Excitations in Twisted Trilayer Optical Lattices}
\title{Ground State Phases and Topological Excitations of Spin-1 Bose-Einstein Condensate in Twisted Optical Lattices}

\author{Tian-Tian Li}
\thanks{These authors contributed equally to this work.}
\affiliation{Key Laboratory of Atomic and Subatomic Structure and Quantum Control (Ministry of Education), Guangdong Basic Research Center of Excellence for Structure and Fundamental Interactions of Matter, School of Physics, South China Normal University, Guangzhou 510006, China}

\author{Ze-Hong Guo}
\thanks{These authors contributed equally to this work.}
\affiliation{Key Laboratory of Atomic and Subatomic Structure and Quantum Control (Ministry of Education), Guangdong Basic Research Center of Excellence for Structure and Fundamental Interactions of Matter, School of Physics, South China Normal University, Guangzhou 510006, China}

\author{Xiao-Ning Wang}
\affiliation{Key Laboratory of Atomic and Subatomic Structure and Quantum Control (Ministry of Education), Guangdong Basic Research Center of Excellence for Structure and Fundamental Interactions of Matter, School of Physics, South China Normal University, Guangzhou 510006, China}

\author{Qizhong Zhu}
\email{qzzhu@m.scnu.edu.cn}
\affiliation{Key Laboratory of Atomic and Subatomic Structure and Quantum Control (Ministry of Education), Guangdong Basic Research Center of Excellence for Structure and Fundamental Interactions of Matter, School of Physics, South China Normal University, Guangzhou 510006, China}
\affiliation{Guangdong Provincial Key Laboratory of Quantum Engineering and Quantum Materials, Guangdong-Hong Kong Joint Laboratory of Quantum Matter, Frontier Research Institute for Physics, South China Normal University, Guangzhou 510006, China}

\date{\today}

\begin{abstract}
Recently, the simulation of moir\'e physics using cold atom platforms has gained significant attention. These platforms provide an opportunity to explore novel aspects of moir\'e physics that go beyond the limits of traditional condensed matter systems. 
Building on recent experimental advancements in creating twisted bilayer spin-dependent optical lattices for pseudospin-1/2 Bose gases, we extend this concept to a trilayer optical lattice for spin-1 Bose gases. Unlike conventional moir\'e patterns, which are typically induced by interlayer tunneling or interspin coupling, the moir\'e pattern in this trilayer system arises from inter-species atomic interactions. 
We investigate the ground state of Bose-Einstein condensates loaded in this spin-1 twisted optical lattice under both ferromagnetic and antiferromagnetic interactions. We find that the ground state forms a periodic pattern of distinct phases in the homogeneous case, including ferromagnetic, antiferromagnetic, polar, and broken axial symmetry phases. Additionally, by quenching the optical lattice potential strength, we examine the quench dynamics of the system above the ground state and observe the emergence of topological excitations such as vortex pairs. 
This study provides a pathway for exploring the rich physics of spin-1 twisted optical lattices and expands our understanding of moir\'e systems in synthetic quantum platforms.
\end{abstract}

\maketitle

\section{introduction}
The recent groundbreaking discovery of unconventional superconductivity \cite{cao2018} and correlated insulator behavior \cite{caoCorrelatedInsulatorBehaviour2018a} in twisted bilayer graphene has generated significant interest in the field of moir\'e physics. Twisted bilayer van der Waals materials, which form long-period moir\'e superlattices, have become a powerful tool for tuning the electronic properties of two-dimensional (2D) quantum materials \cite{bistritzer2011}. Among the most notable electronic features is the emergence of moir\'e flat bands, providing a platform for studying strongly correlated physics \cite{chen2019, codecido2019, sharpe2019, kerelsky2019, wilson2021, li2021}. Another major breakthrough is the realization of topological models that do not rely on external magnetic fields \cite{yu2019giant, wu2019, devakul2021, zeng2023, xu2023, li2024}. An especially exciting achievement that combines flat-band physics and topology is the realization of the fractional quantum anomalous Hall effect without the need for an external magnetic field \cite{Liu2023a, zeng2023, xu2023, cai2023nature}. Moir\'e materials have become an essential platform in condensed matter physics, enabling the study and quantum simulation of exotic phases that were previously nearly impossible to explore.

Interlayer coupling, essential for the formation of moir\'e patterns, along with electron interaction strength in materials, is typically challenging to adjust in condensed matter systems. Consequently, there has been growing interest in exploring alternative platforms for studying moir\'e physics. Promising candidates for realizing moir\'e-related phenomena include optical systems \cite{huang2016, wang2020, fu2020, arkhipova2023}, acoustic systems \cite{wu2022a, duan2023, oudich2024}, and cold atom systems \cite{soltan-panahi2011a, jo2012, luo2021, Lee2022, meng2023, gonzalez-tudela2019, madronero2024, zeng2024, GaoChao2024, tian2024}. 

The study of moir\'e physics in ultracold atom systems offers significant advantages due to the high degree of controllability these systems provide. A key breakthrough has been the experimental realization of twisted bilayer square optical lattices \cite{meng2023}. There are also theoretical proposals to create twisted bilayer hexagonal lattices \cite{luo2021,Lee2022} and twisted three-dimensional optical lattices \cite{GaoChao2024}. More recently, researchers have proposed a new mechanism for moir\'e lattice formation \cite{zeng2024,tian2024}, which relies on interparticle interactions rather than interlayer tunneling.

Moir\'e physics in cold atomic systems has primarily focused on twisted bilayer setups, which correspond to pseudospin-1/2 systems. A natural extension of this idea is to explore twisted trilayer optical lattices, which are associated with spin-1 systems. In this work, we investigate this possibility by studying a spin-1 Bose-Einstein condensate (BEC) loaded into twisted trilayer optical lattices. 
Our motivation is twofold. First, in homogeneous space, spin-1 BECs are known to exhibit a much richer ground-state phase diagram compared to pseudospin-1/2 systems, with four possible phases: the ferromagnetic (FM) phase, antiferromagnetic (AFM) phase, polar (P) phase, and the broken axial symmetry (BA) phase, depending on spin-dependent interactions and external magnetic fields \cite{ho1998, ohmi1998, stenger1998}. This diverse phase diagram is expected to enable more complex and intriguing physics when combined with the spatial inhomogeneity induced by a moir\'e lattice. Second, twisted trilayer 2D materials \cite{park2021,zhu2020,bai2023,yao2024topological}, which have attracted significant interest, could potentially be simulated using twisted spin-1 atoms in spin-dependent optical lattices. 
Additionally, for a ferromagnetic spin-1 BEC, quenching the magnetic field across the phase boundary from the P phase to the BA phase can excite the formation of topological defects. The relationship between defect density and quench speed can be understood through the Kibble-Zurek mechanism \cite{zurek1985, Damski2005, Damski2007, Anquez2016}. In untwisted optical lattices, the lattice potential acts as a scalar potential and does not interact with the internal spin degrees of freedom. However, in twisted optical lattices, the spin-dependent lattice potentials act as spatially varying magnetic fields, generating intriguing spatial patterns in spin configurations. 
These possibilities motivate our investigation of spin-1 BECs in twisted trilayer optical lattices.

Here, in this paper, we study in detail the properties of spin-1 BEC loaded in a twisted trilayer optical lattices, focusing on the ground state phases and topological excitations following a quench in the lattice potential strength. 
We find that the ground state consists of periodic patterns of different phases in the homogeneous case. For both ferromagnetic and antiferromagnetic interactions, all homogeneous phases including FM, AFM, P and BA phases can appear in the local phase pattern under certain conditions. This spatially dependent phase pattern demonstrates moir\'e lattice periodicity. After quenching the lattice potential, vortex pairs are excited above the ground state, and their spatial distribution also displays the characteristic moir\'e periodicity. For the system we consider here, since interspin coupling is not included at the single-particle level, the formation of the moir\'e pattern is entirely due to atomic interactions.

The structure of this paper is organized as follows. Section \ref{sec2} introduces the model of a spin-1 BEC in twisted optical potentials and formulates the system using the Gross-Pitaevskii (GP) equations within a mean-field theory framework. 
In Section \ref{sec3}, we briefly review the classification of ground state phases in homogeneous systems. Then, by numerically solving the GP equations, we demonstrate the distribution of ground state phases for the spin-1 BEC in a spatially dependent twisted optical lattice potential. We classify the local phases by comparing them to the ground state phases in homogeneous systems and analyze how their spatial distribution changes with varying lattice potential strength.
Section \ref{sec4} presents numerical simulations of the dynamics of a quenched spin-1 BEC in a twisted optical lattice, highlighting the emergence of topological defects. Finally, Section \ref{sec5} summarizes the main findings of this study.

\section{System and model Hamiltonian}
\label{sec2}
\begin{figure}[tbp] \centering
  \includegraphics[width=0.99\linewidth]{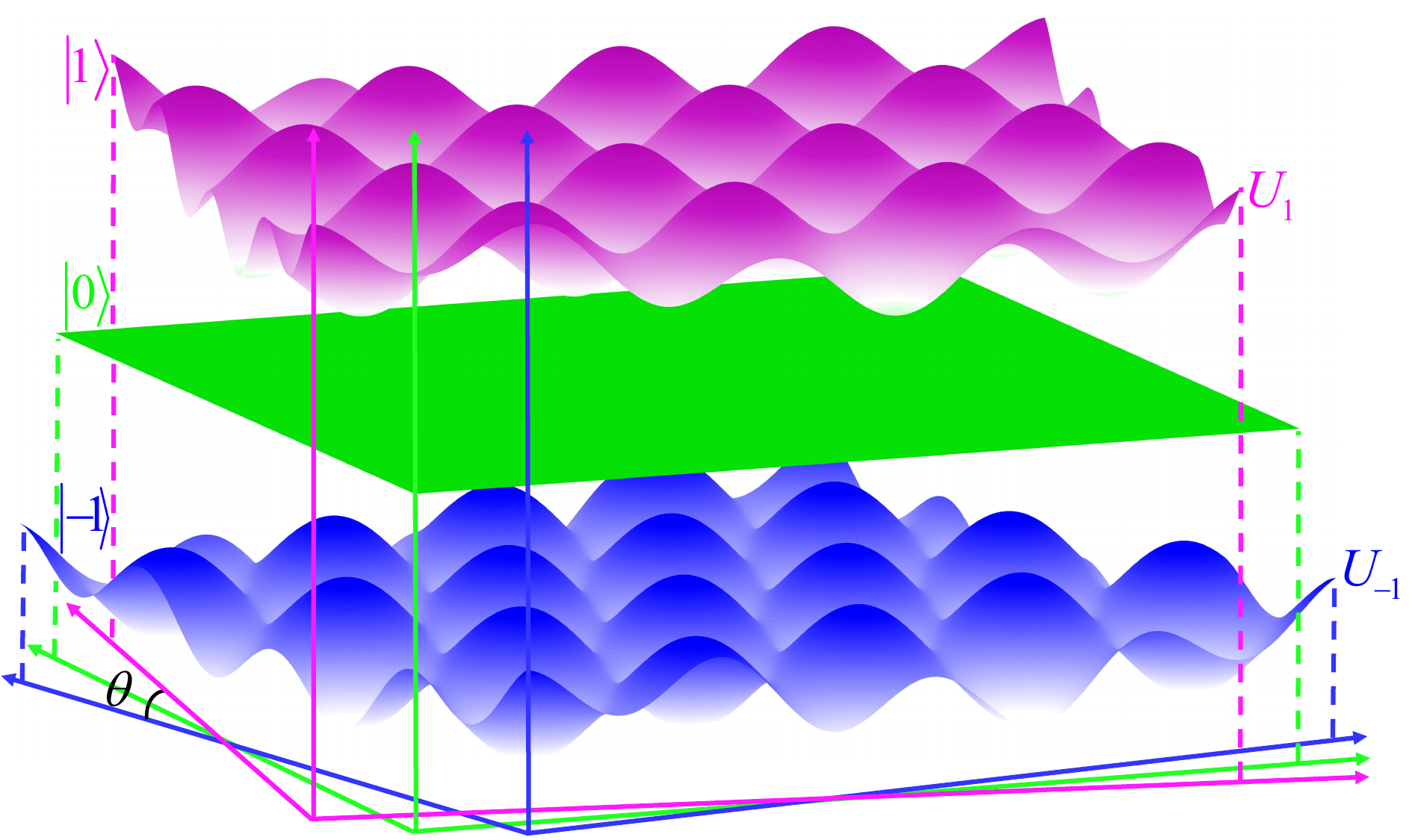}
  \caption{Lattice potential experienced by spin-1 BEC (the overall two-dimensional trapping potential is not shown). 
  Two sets of square lattices $U_1$ (purple) and $U_{-1}$ (blue) with a relative angle $\theta$ on the horizontal plane form a spin-dependent lattice potential. 
  Atoms in the spin states $\ket{F = 1, m_F = 1}$ and $\ket{F = 1, m_F = -1}$ are respectively loaded in the lattice potentials $U_1$ and $U_{-1}$, while atoms in the spin state $\ket{F = 1, m_F = 0}$ do not experience any lattice potential. There is no interspin coupling at single-particle level.}
  \label{fig1}
\end{figure}

We consider a spin-1 BEC consisting of three components, corresponding to the three hyperfine states $\ket{F=1,m_F=-1}$, $\ket{F=1,m_F=0}$, and $\ket{F=1,m_F=1}$. The BEC is loaded into a twisted optical lattice, where the $m=1$ and $m=-1$ components experience square optical lattice potentials that are rotated relative to each other by an angle $\theta$. The $m=0$ component, however, does not experience the optical lattice potential. Specifically, the lattice potentials for the $m=\pm1$ components are rotated by angles of $\pm\theta/2$ with respect to the lattice-free $m=0$ component. All three components are confined within a harmonic trap. Additionally, the strong confinement along the $z$-axis freezes motion in that direction, making the system quasi-2D.
The square lattice potentials experienced by the $m=1$ and $m=-1$ components are denoted by $U_1$ and $U_{-1}$, respectively, as illustrated in Fig. \ref{fig1}. These two potentials can be experimentally generated using techniques similar to those used in twisted bilayer optical lattice setups \cite{meng2023}. It is important to note that there is no externally imposed interspin coupling, such as spin-flip processes, in this system. As a result, in the absence of atomic interactions, these three components behave as completely independent species. Consequently, the observed moir\'e pattern arises purely from atomic interactions between the components.

Within the mean-field theory framework, the energy functional of this system in 2D reads,
\begin{equation}\label{functional}
\begin{split}
\mathcal{E}=&\int d^2 \mathbf{r}\biggl\{\sum_{m=-1}^{1}\psi^{*}_{m}\left[-\frac{\hbar^{2}\nabla^{2}}{2M}+V_{\mathrm{trap}}(\mathbf{r})+U_m(\mathbf{r})\right]\psi_{m}\\
&+\frac{C_0}{2}{n}^2+\frac{C_1}{2}\left| \mathbf{f}\right|^2\biggr\},
\end{split}
\end{equation}
where $\psi_m$ $(m=0, \pm1)$ is the three-component spinor wave function of the atoms condensed in spin state $\ket{F=1, m_F=m}$, $M$ is the atomic mass and $V_{\mathrm{trap}}(\mathbf{r})=M\omega^2(x^2+y^2)/2$ is the 2D harmonic trap. 
The 2D interaction strengths are $C_0=\sqrt{8\pi}\hbar^2{\left(a_0+2a_2\right)}/{3Ml_z}$ and $C_1=\sqrt{8\pi}\hbar^2{(a_{2}-a_{0})}/{3Ml_z}$, where $l_z=\sqrt{\hbar/M\omega_z}$ denotes the characteristic length along the $z$ axis \cite{2Dtrap2000}, and $a_0$ and $a_2$ are 3D $s$-wave scattering lengths in the total spin-0 and spin-2 channels, respectively. $C_1>0$ corresponds to antiferromagnetic interaction and $C_1<0$ corresponds to ferromagnetic interaction.
The 2D effective density $n(\mathbf{r})=\sum_{m}\left| \psi_m\right| ^2$ satisfies the constraint $\int d^2 \mathbf{r} n\left(\mathbf{r}\right)=N$ with $N$ being the total atom number.
The spin density vector is $\mathbf{f} =\left(f_x, f_y, f_z\right)=(\psi^\dagger\hat{F_x}\psi$, $\psi^\dagger\hat{F_y}\psi$, $\psi^\dagger\hat{F_z}\psi)$, where $\hat{F_x}$, $\hat{F_y}$ and $\hat{F_z}$ are the $3\times3$ Pauli matrices for spin-1 representation. 
The spin-dependent optical lattice $U_m(\mathbf{r})$ in Eq. (\ref{functional}) takes the following form
\begin{align}\label{potentials}
U_{0}(\mathbf{r})=&0,\\
U_{1}(\mathbf{r})=&V_{0}[\sin^{2}(kx\cos{\frac{\theta}{2}}-ky\sin{\frac{\theta}{2}})\nonumber\\
&+\sin^{2}(ky\cos{\frac{\theta}{2}}+kx\sin{\frac{\theta}{2}})]-V_1,\\ 
U_{-1}(\mathbf{r})=&V_{0}[\sin^{2}(kx\cos{\frac{\theta}{2}}+ky\sin{\frac{\theta}{2}})\nonumber\\
&+\sin^{2}(ky\cos{\frac{\theta}{2}}-kx\sin{\frac{\theta}{2}})]-V_2,
\end{align}	
where $k ={2\pi}/{\lambda}$ is the wave number of the laser field, and $\lambda$ is the wavelength of the laser. The optical lattice potentials experienced by $m=\pm1$ component are rotated by $\pm\theta/2$ respectively and the relative twist angle between these two components is $\theta$. $V_0$ and $V_i$ $\left(i=1,2\right)$ respectively denote the lattice potential depth and the detuning in unit of the recoil energy of the optical lattice $E_r = {\hbar^2k^2}/{2M}$. $V_i$ $\left(i=1,2\right)$ can be tuned by external magnetic field through the linear and quadratic Zeeman energy. By comparing the lattice potential energy with the energy of a spin-1 BEC in a magnetic field, expressed as $E(\mathbf{r})=\sum_m U_m(\mathbf{r})|\psi_m|^2=\sum_m (-P(\mathbf{r})m+Q(\mathbf{r})m^2)|\psi_m|^2$, which includes both linear and quadratic Zeeman terms, we can define the coefficients $P(\mathbf{r}) =({U_{-1}(\mathbf{r})-U_1(\mathbf{r})})/{2}$ and $Q(\mathbf{r}) =({U_{-1}(\mathbf{r})+U_1(\mathbf{r})})/{2}$. Here, $P(\mathbf{r})$ and $Q(\mathbf{r})$ represent the coefficients of the effective linear Zeeman energy and quadratic Zeeman energy, respectively. Both coefficients are spatially dependent and directly connected to the twisted optical lattice potentials. As shown in Fig. \ref{fig12}, the coefficients $P(\mathbf{r})$ and $Q(\mathbf{r})$ are spatially varying within a moir\'e period.
In the non-interacting case, different components are decoupled, so the ground state of the system does not have a moir\'e structure. However, when atomic interactions are introduced, the twisted lattice potential experienced by the $m=\pm1$ components is transferred to the $m=0$ component, leading to the emergence of an interaction-induced moir\'e pattern.

\begin{figure}[tbp] \centering
	\includegraphics[width=0.99\linewidth]{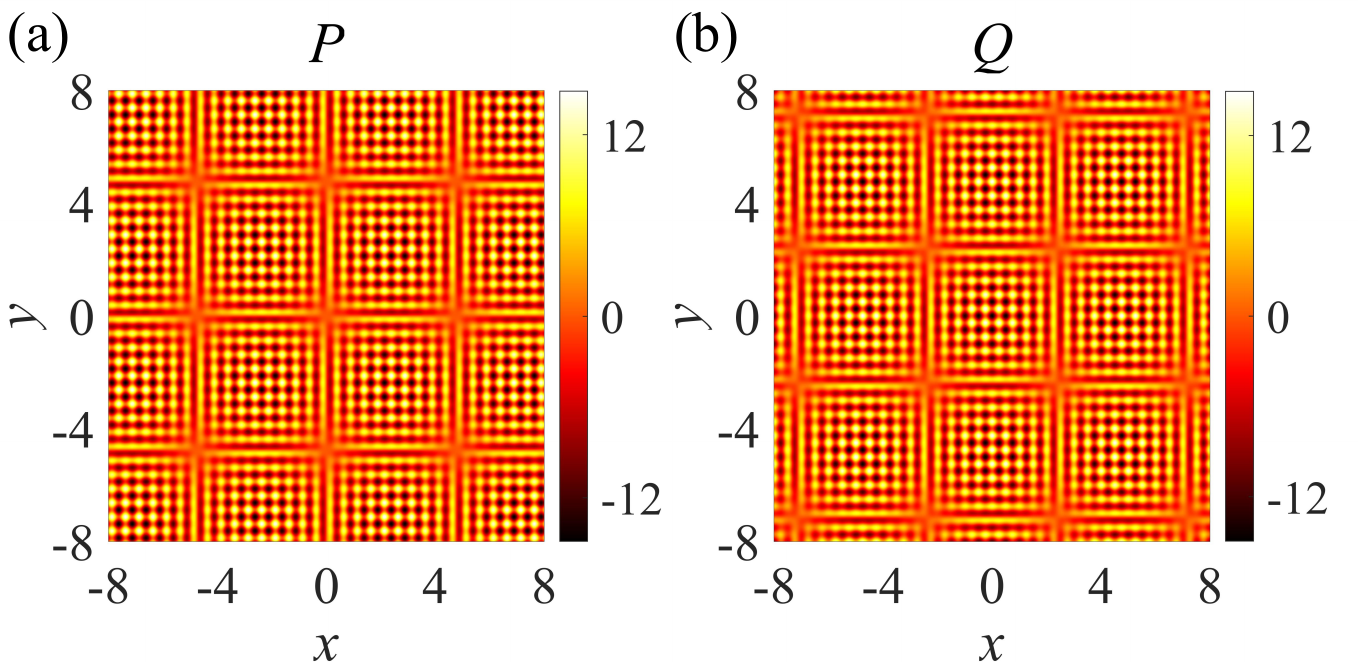}
	\caption{The spatial variation of the coefficients of effective linear ($P(\mathbf{r})$) and quadratic ($Q(\mathbf{r})$) Zeeman energy at twisted angle $\theta=\pi/30$ is shown in (a) and (b), respectively. The length unit is $\lambda$. The lattice depths are $V_0=V_1=V_2=15.0 E_r$.}
	\label{fig2}
\end{figure}

The dynamics of the spinor wave functions obey the multi-component time-dependent GP equations 
\begin{equation} \label{DGPE}
  \begin{split}
    i\hbar\frac{\partial \psi_1}{\partial t}&=[-\frac{\hbar^2\nabla^2}{2M}+V_{\mathrm{trap}}(\mathbf{r})+U_1(\mathbf{r})+C_0 n+C_1f_z]\psi_{1}\\
    &+\frac{C_1}{\sqrt{2}}f_-\psi_{0},\\
    i\hbar\frac{\partial \psi_0}{\partial t}&=[-\frac{\hbar^2\nabla^2}{2M}+V_{\mathrm{trap}}(\mathbf{r})+U_{0}(\mathbf{r})+C_{0}n]\psi_{0}+\frac{C_{1}}{\sqrt{2}}f_{+}\psi_{1}\\
    &+\frac{C_{1}}{\sqrt{2}}f_{-}\psi_{-1},\\
    i\hbar\frac{\partial \psi_{-1}}{\partial t}&=[-\frac{\hbar^2\nabla^2}{2M}+V_{\mathrm{trap}}(\mathbf{r})+U_{-1}(\mathbf{r})+C_{0}n-C_1f_z]\psi_{-1}\\
    &+\frac{C_{1}}{\sqrt{2}}f_{+}\psi_{0},	
  \end{split}
\end{equation}
where $f_{\pm}=f_x\pm if_y$. The real-time dynamics and ground state of the system can be determined by solving the time-dependent GP equations in the real-time and imaginary-time domains, respectively. 

\section{ground state phases in spin-1 moir\'e system}
\label{sec3}
In the following, we study the ground state properties of spin-1 BEC within spatially varying twisted optical lattice potentials, focusing on the spatial distribution of local phase pattern.

\begin{figure*}[tb] \centering
  \includegraphics[width=0.99\textwidth]{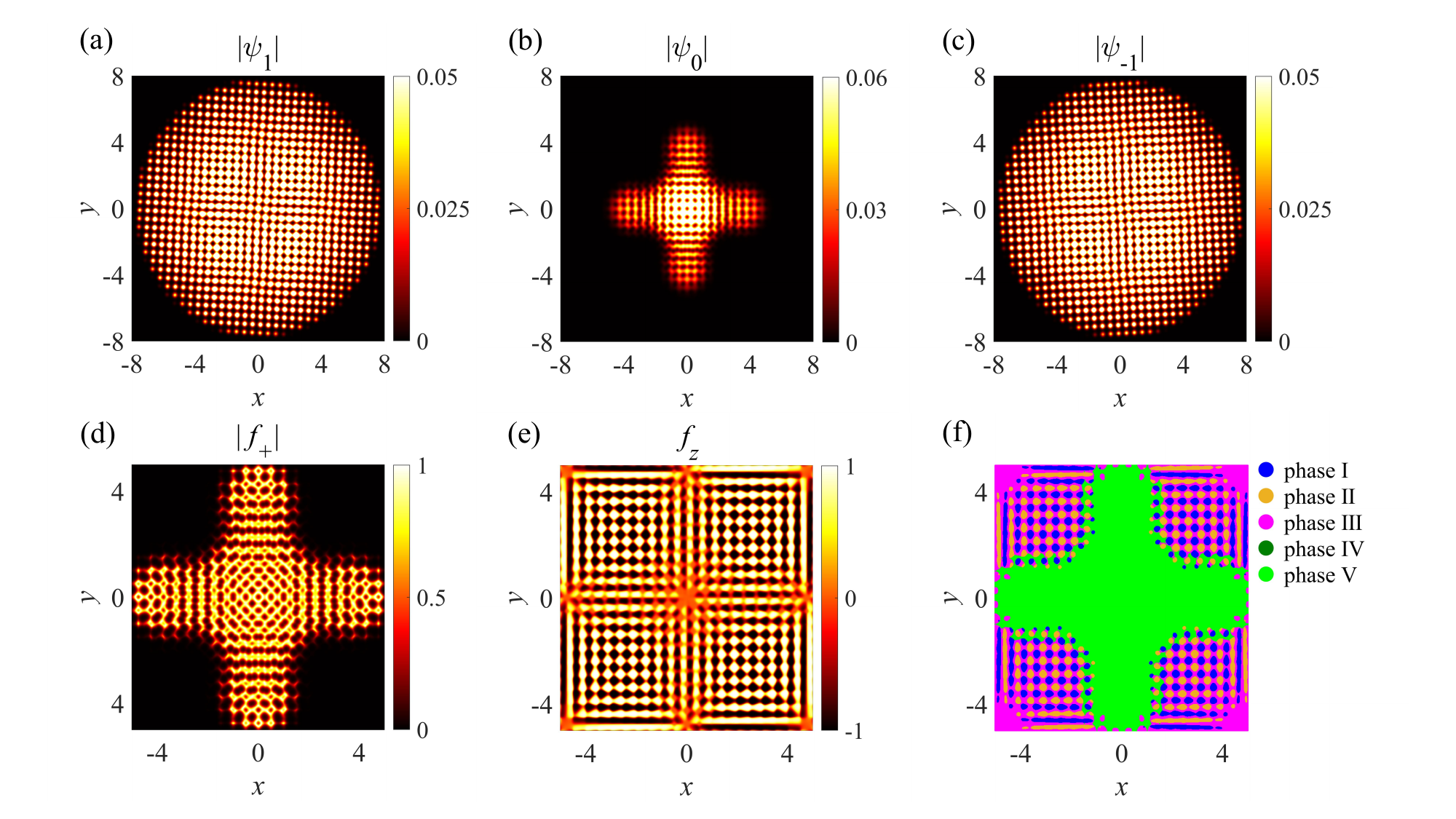}
  \caption{(Color online) The ground state properties of antiferromagnetic spin-1 BEC in twisted optical lattice potentials. 
  (a)-(c) illustrate the ground state spatial density distribution of spin-1 BEC for the
  $m=1$, $m=0$ and $m=-1$ components, respectively. 
  The density distributions for all components form moir\'e patterns.
  (d)-(e) show the spatial distribution of transverse magnetization $\left| f_+\right|$ and longitudinal magnetization $f_z$.
  (f) presents the spatial distribution of local phases at the ground state for a threshold value of $\epsilon = \delta = 0.05$. 
  Phases I-V correspond to FM phase for $m=1$, FM phase for $m=-1$, AFM phase, P phase, and BA phases, respectively.
  The other parameters are $w \approx 2\pi \times410$Hz, $C_0n=1.936E_r$, $C_1n=0.007E_r$, $V_0=15.0E_r$, $V_1=V_2=12.5E_r$, $\theta=\pi/30$.}
  \label{fig3}
\end{figure*}

\subsection{Ground state phases in homogeneous system}
\label{subsec3-1}
Before presenting the details of the results, let us briefly review the ground state phases for a spin-1 ferromagnetic and antiferromagnetic BEC in the homogeneous case, with the presence of external magnetic field. Depending on the sign of interaction strength $C_1$ and the linear and quadratic Zeeman energy in the presence of external magnetic field, there are four distinct types of phases in total \cite{kawaguchi2012}: 

$\left(1\right)$ FM phase. Atoms in this phase are fully magnetized along the positive or negative spin $z$ axis. The wave function of the former (later) state is only nonvanishing in the $m=1$ ($m=-1$) component, i.e., $\psi=\sqrt{N}(\psi_1,0,0)$ ($\psi=\sqrt{N}(0,0,\psi_{-1}$)), with the longitudinal magnetization $f_z=\left| \psi_1\right| ^2-\left| \psi_{-1}\right| ^2=1$ ($f_z=-1$) and transverse magnetization $\left| f_{+}\right|=0$.

$\left(2\right)$ AFM phase. Atoms in this phase are distributed in $m=1$ and $m=-1$ components, i.e., $\psi=\sqrt{N}(\psi_1,0,\psi_{-1})$ with the longitudinal magnetization $f_z \in (-1,1)$ and transverse magnetization $\left| f_{+}\right|=0$. The magnetization direction is determined by the coefficient of linear Zeeman energy.

$\left(3\right)$ P phase. Wave functions of this phase only have the $m=0$ component, i.e., $\psi=\sqrt{N}(0,\psi_0,0)$ and the magnetization in all directions vanish in this phase.

$\left(4\right)$ BA phase. Due to the spontaneous breaking of the axial rotational symmetry, the atoms in this phase have nonvanishing transverse magnetization, i.e., $\left| f_{+}\right| >0$, indicating that the wave function have nonzero value in all three components. In this phase, the longitudinal magnetization $f_z \in (-1,1)$, and the transverse magnetization $\left| f_{+}\right| \in (0,1)$.

The four phases are determined by minimizing the total energy, which includes the atomic interaction energy as well as the linear and quadratic Zeeman terms, in a uniform system. The kinetic energy is neglected, because in a uniform system, it is always zero and is therefore not considered in the minimization process. In contrast, in an inhomogeneous system, the kinetic energy is inevitably nonzero. As a result, minimizing the total energy in such a system, including the kinetic energy, can lead to new phases with local wave functions or order parameters that differ from the four phases mentioned above, as will be explained below.

\subsection{Ground state properties and local phase patterns in inhomogeneous moir\'e system}
\label{subsec3-2}
We now present the local phase patterns of the ground state for a spin-1 BEC in twisted optical lattices. By solving Eq. \ref{DGPE} through imaginary time evolution, we first examine the ground state properties in the case of antiferromagnetic interactions ($C_1 > 0$). The results for ferromagnetic interaction ($C_1 < 0$) are similar and provided in the Appendix. 
Numerical simulations of the density distribution reveal that atoms in the spin states $m=1$ and $m=-1$, confined within the 2D twisted optical lattice, both display moir\'e pattern distributions, as shown in Figs. \ref{fig3}(a) and \ref{fig3}(c), respectively. Even in the absence of single-particle interspin coupling, the atomic interaction term in the GP equation, which causes interspin scattering, mixes the three components and leads to the formation of moir\'e patterns also in the $m=0$ component, as illustrated in Fig. \ref{fig3}(b). This result supports the concept of interaction-induced moir\'e patterns \cite{zeng2024, tian2024}, which represents a novel mechanism for moir\'e lattice formation. This mechanism differs from the conventional approach, where single-particle interlayer coupling (analogous to interspin coupling here) is responsible for creating such patterns.

The spatial variation of different local phases in the ground state can be understood using the local density approximation, which is expected to be accurate enough for small twist angles. In this case, the moir\'e period is long, and the spatial variation of each term in the Hamiltonian is smooth. Consequently, using the spatial maps of $P(\mathbf{r})$ and $Q(\mathbf{r})$, each local region can be treated as approximately uniform but with distinct values of $P(\mathbf{r})$ and $Q(\mathbf{r})$. The ground state should closely resemble the uniform ground state while accounting for spatial variation of $P(\mathbf{r})$ and $Q(\mathbf{r})$. This approximation forms the foundation of the continuum model widely used to study twisted bilayer 2D materials \cite{wu2019,yu2019giant,bistritzer2011} and plays a crucial role in the concept of the topological mosaic pattern \cite{tong2017}. However, at finite twist angles, this variation incurs a kinetic energy cost. As a result, the ground state obtained by minimizing the total energy, including the kinetic energy, exhibits deviations from a simple repetition of the uniform ground state across different spatial regions.

Similar to the homogeneous case, where different phases are characterized by distinct order parameters, we analyze the ground state phases of the spin-1 BEC in the inhomogeneous moir\'e lattice by examining the transverse magnetization $|f_+|$ (Fig. \ref{fig3}(d)) and the longitudinal magnetization $f_z$ (Fig. \ref{fig3}(e)), for the parameters used in Figs. \ref{fig3}(a)-(c). The results clearly demonstrate that the magnetization properties of the atoms exhibit spatial dependence, with the spontaneous emergence of a moir\'e pattern distribution. From the alternating distribution of $f_z$ and the presence of large regions where $|f_+|$ remains nonzero, it is evident that the ground state contains diverse local phases, each characterized by different local order parameters.

\begin{figure*}[tb] \centering
  \includegraphics[width=0.93\linewidth]{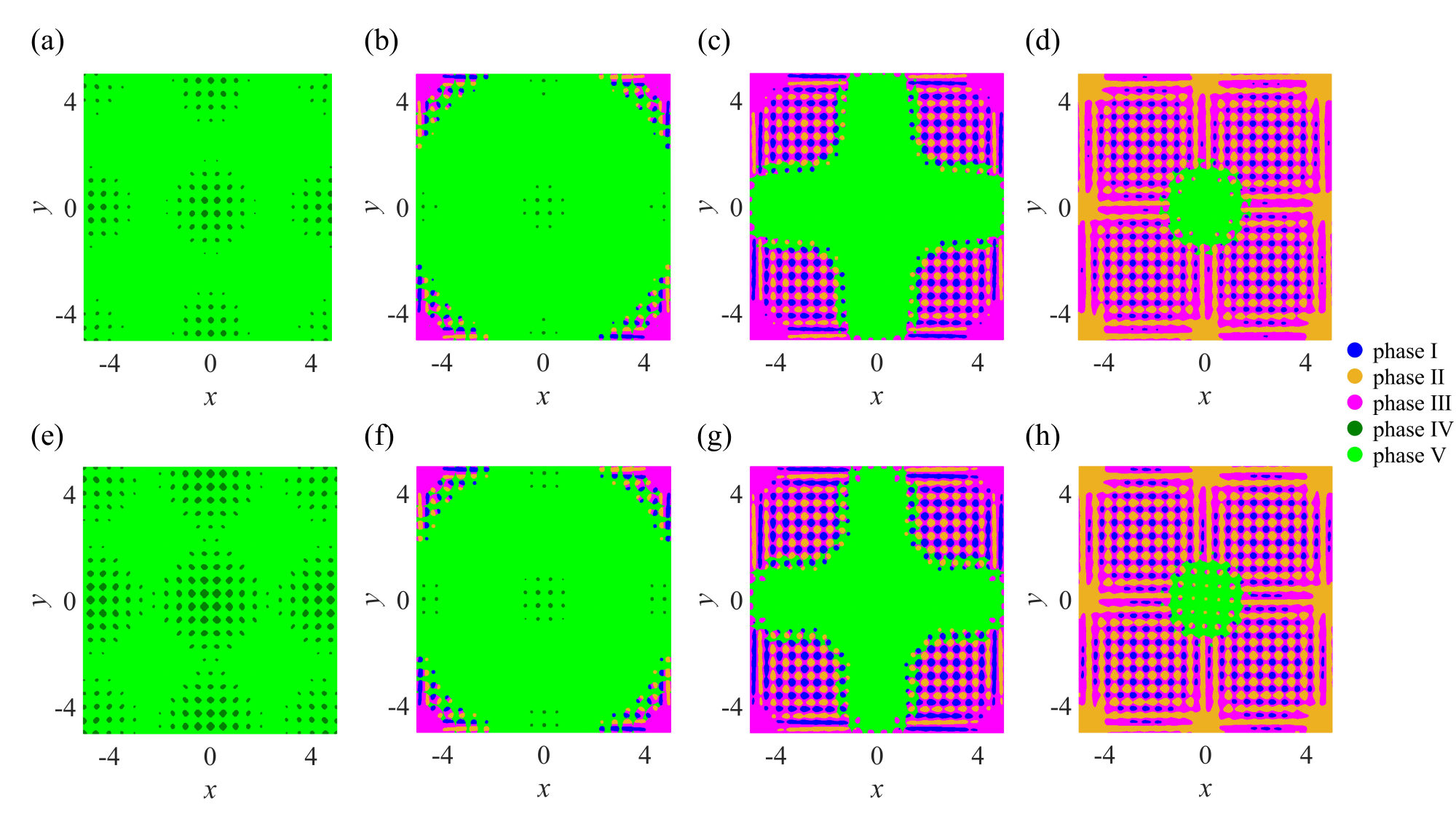}
  \caption{(Color online) The spatial distribution of local phases for antiferromagnetic spin-1 BEC under different lattice depths and numerical thresholds. The lattice potential depth is $V_1=V_2=7.5E_r$ in (a) and $V_1=V_2=9.0 E_r$ in (b). From (c) to (d), $V_1=12.5E_r$ and $V_2=12.5E_r$, $14.5E_r$, respectively. The top and bottom rows correspond to numerical thresholds $\epsilon=\delta=0.03$ and $\epsilon=\delta=0.05$, respectively. The other parameters are $w \approx 2\pi \times410$Hz, $C_0n=1.936 E_r, C_1n=0.007 E_r, V_0=15.0 E_r$, $\theta=\pi/30$. The definitions of phases I-V are the same as those in Fig. \ref{fig3}.
  }
  \label{fig4}
\end{figure*}

To be specific, by analyzing the magnetization distribution characteristics of each local phase, we classify the ground state of the spin-1 BEC in the moir\'e lattice under antiferromagnetic interactions. Unlike the uniform case, where only the $m = \pm 1$ components of the wave function are nonzero in the FM phase, the $m = 0$ component is zero in the AFM phase, and only the $m = 0$ component is nonzero in the P phase, the inhomogeneous moir\'e lattice generally has all three components of the wave function nonzero. Therefore, to compare with the homogeneous case, the classification criteria must be adjusted to properly characterize local phases in the inhomogeneous system.

We set the thresholds $\epsilon$ and $\delta$ to define the ranges within which the transverse and longitudinal magnetization deviate from the homogeneous case, respectively. For instance, the thresholds can be chosen as $\epsilon = \delta = 0.05\ll 1$. The detailed classification process is as follows: In numerical simulations, a phase with transverse magnetization $\left| f_+ \right| \geq \epsilon$ is classified as the BA phase. When $\left| f_+ \right| < \epsilon$, the longitudinal magnetization $f_z$ is used to further classify the phase. Specifically, if the longitudinal magnetization satisfies $|f_z| \in [1-\delta, 1]$, the phase is classified as the FM phase, where atoms are distributed primarily in the $m = 1$ or $m = -1$ component. If $|f_z| \in (\delta, 1-\delta)$, it is identified as the AFM phase. Finally, when $|f_z| \in [0, \delta]$, the system can be either in the AFM state or the P phase. In this case, the characteristics of the wave function must be analyzed further. If the wave function's major value is in the $m = \pm 1$ components, it is classified as the AFM phase. Otherwise, if the major value is in the $m = 0$ component, it is identified as the P phase.
This classification approach cannot always unambiguously distinguish certain phases, as the result depends on the chosen thresholds. We have tested thresholds of 0.03 and 0.05, finding that the overall phase patterns remain consistent.

With a numerical threshold of $\epsilon = \delta = 0.05$, the spatial distribution of the system's ground state phases is shown in Fig. \ref{fig3}(f).  
By examining Fig. \ref{fig3}(f) and Figs. \ref{fig4}(a)-(d), one observes four distinct local phases for certain parameters, where phases I and II represent FM phases with opposite magnetic polarizations. In other words, all the phases found in the homogeneous system also appear in the moir\'e inhomogeneous system under specific conditions. The BA phase, absent in the homogeneous system with antiferromagnetic interaction ($C_1 > 0$), emerges in this inhomogeneous system. Its emergence reflects the spontaneous breaking of axial rotational symmetry, associated with a fixed phase angle in $f_+$, as the ground state energy is invariant with respect to this phase angle. 
The appearance of BA phase, absent in the homogeneous system, is attributed to the inclusion of the kinetic energy term in the total energy minimization, which arises from spatial inhomogeneity. 
Furthermore, the presence of the moiré pattern in the ground state can be explained by the spatial variations of $P(\mathbf{r})$ and $Q(\mathbf{r})$.

\begin{figure*}[tb] \centering
  \centering
  \includegraphics[width=0.99\linewidth]{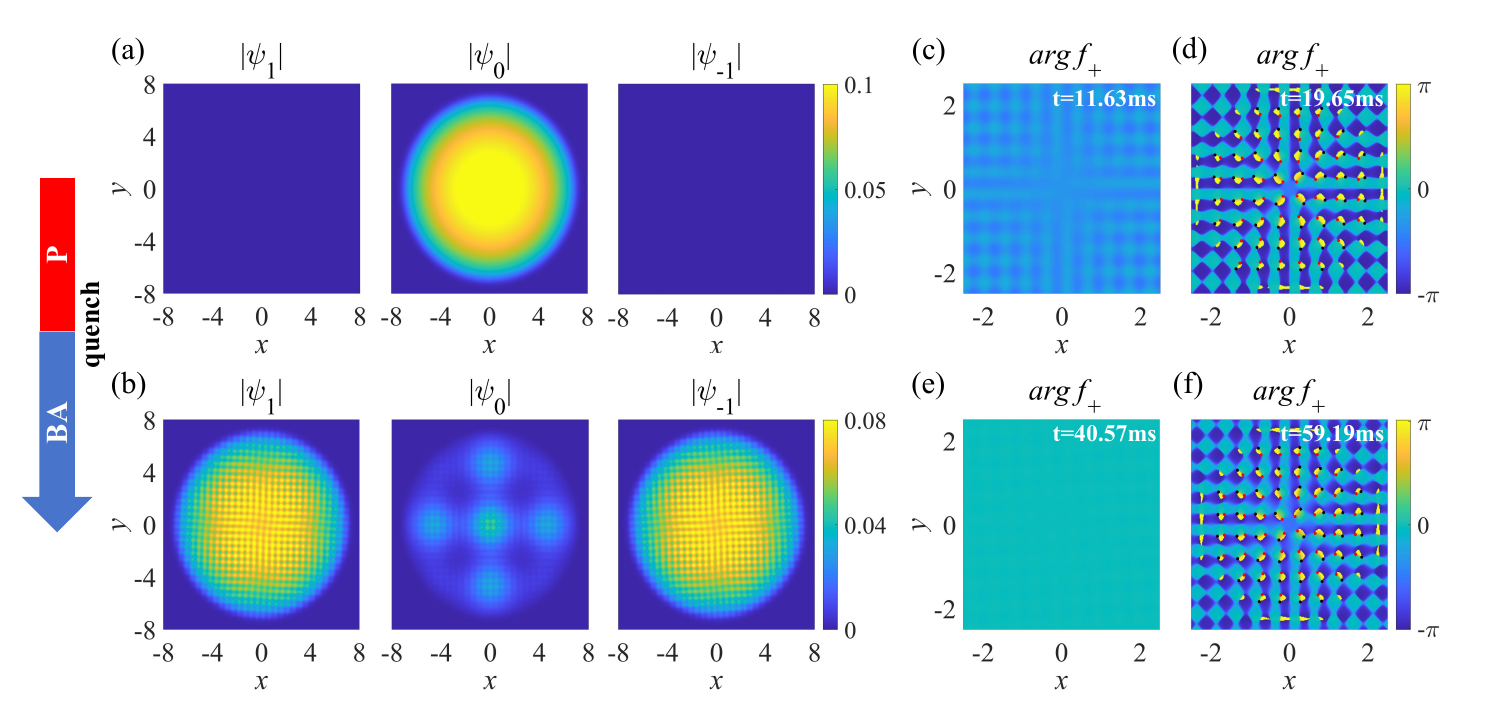}
  \caption{(Color online) Evolution of wave functions and transverse magnetization of antiferromagnetic spin-1 BEC in the twisted optical lattice potentials following the instantaneous quench from $V_0=1.5E_r$ to $V_0=1.15E_r$ (shown by the arrow). The ground state primarily consists of P and BA phases under the given parameters before and after the quench, respectively. (a)-(b) illustrate the evolution of the density distribution before and after the quench, respectively. (c)-(f) show the time evolution of the phase angle of $f_+$ ($arg f_+$) at different moments. The other parameters are $w \approx 2\pi \times360$Hz, $C_0n=1.936E_r$, $C_1n=0.007E_r$, $\theta=\pi/30$, and $V_1=V_2=1.11E_r$.
  }\label{fig5}
\end{figure*}

Furthermore, we discuss the impact of twisted optical lattice strength and numerical thresholds on the spatial distribution of local phases at ground state. In Fig. \ref{fig4}, we show how the classification of various local phases depends on different lattice potential strengths and numerical thresholds. Figs. \ref{fig4}(a)-(d) illustrate the effect of lattice potential strengths on local phase distribution, which can be understood in terms of the effective linear and quadratic Zeeman coefficients, $P(\mathbf{r})$ and $Q(\mathbf{r})$, determined by the external lattice potential. Specifically, when $V_1$ is fixed and $V_2$ increases, both $P(\mathbf{r})$ and $Q(\mathbf{r})$ decrease, which favors the emergence of the AFM phase and reduces the size of the BA phase. This trend is evident in Figs. \ref{fig4}(a)-(d) and (e)-(h). We also examine the effect of using a different numerical threshold and present the results in Figs. \ref{fig4}(e)-(h). The comparison shows that the two thresholds, 0.03 and 0.05, result in very similar local phase distributions. This indicates that the phase classification described above is relatively robust against the choice of numerical thresholds.

\section{Topological excitations in inhomogeneous moir\'e system}
\label{sec4}
In a homogeneous system, it is known that quenching the magnetic field across the phase boundary from the symmetry unbroken phase to symmetry breaking phase can generate topological excitations \cite{sadler2006, saito2007}. 
In the twisted optical lattice system, with its intriguing local phase pattern, quenching the effective magnetic field is expected to create even more diverse and complex topological excitations.  

For easier observation of topological excitations, we initially choose parameters such that the entire spatial region is predominantly in the P phase at the beginning. Then, by instantaneously quenching the strength of the twisted lattice potential $V_0$, the system now has a ground state where the BA phase dominates. The density distributions corresponding to the ground states for the initial and final quench parameters are shown in Figs. \ref{fig5}(a) and (b). 
Afterwards, we numerically solve the time evolution of Eq. \ref{DGPE} to investigate the characteristics of the quenching dynamics. Figs. \ref{fig5}(c)-(f) show the phase angle profile of the transverse magnetization after the instantaneous quench. The results reveal that at $t = 19.65$ ms, a large number of vortices spontaneously emerge in regions with more pronounced lattice overlap. Around these vortices, the in-plane spin direction undergoes a rotation of $\pm2\pi$. Specifically, as depicted by the red and black dots in Figs. \ref{fig5}(d) and (f), the red dots indicate vortices with a phase winding of $2\pi$, corresponding to positively charged vortices. Meanwhile, the black dots mark vortices with a phase winding of $-2\pi$, corresponding to negatively charged vortices. Notably, these vortices always form in pairs with total charge being zero, and their spatial distribution reflects the periodicity of the moir\'e lattice. 

As time evolves, the spontaneously generated vortices gradually disappear but reappear at later moments, such as at $t = 59.19$ms. Over longer periods, we have observed that the creation and annihilation of vortex pairs approximately follow periodic oscillations. This phenomenon is closely linked to the spin-mixing dynamics previously studied in homogeneous spin-1 BECs  \cite{widera2005,zhang2005,mahumd2013,dan2013}. During the quenching process, the coherent spin-mixing processes $\hat{\psi}_{1}^{\dagger}\hat{\psi}_{-1}^{\dagger}\hat{\psi}_{0}\hat{\psi}_{0}$ and $\hat{\psi}_{0}^{\dagger}\hat{\psi}_{0}^{\dagger}\hat{\psi}_{1}\hat{\psi}_{-1}$ act as a Rabi-like coupling between the P and BA phases \cite{widera2005,zhang2005,mahumd2013,dan2013,zhu2015}. 
In homogeneous spin-1 BECs, this spin-mixing dynamic has been extensively studied and is known to induce periodic oscillations of spin polarization. In the present system, which is inhomogeneous and features moir\'e periodicity, a similar spin-mixing behavior is expected in local regions, though with varying parameters. However, the global dynamics of the system are far more complex than in the homogeneous case and require further investigation. The periodic creation and annihilation of vortex pairs can be detected using magnetization-sensitive phase-contrast imaging \cite{sadler2006}.

\section{Conclusion and Acknowledgement}
\label{sec5}
In summary, we have proposed a novel twisted trilayer optical lattice system and investigated the ground state properties and topological excitations of a spin-1 BEC within such lattices. 
Using numerical solutions of the Gross-Pitaevskii equations, we studied the spatial distribution of local phases at the ground state in this inhomogeneous moir\'e system. Our results show that an interaction-induced moir\'e pattern can spontaneously form in this twisted spin-1 BEC system, even in the absence of single-particle coupling between different spin states. 
We found that by tuning the parameters of the twisted optical lattice potential, all ground state phases observed in homogeneous systems can also appear in the inhomogeneous moir\'e system. Additionally, due to the spatial inhomogeneity, new phases emerge that are absent in homogeneous cases, whether the spin-1 BEC is ferromagnetic or antiferromagnetic. 
Furthermore, we quenched the optical lattice depth, transitioning from parameters where the P phase dominates to parameters where the BA phase is dominant. During this process, we observed the emergence of topological excitations, i.e., vortex pairs. Given the experimental realization of twisted spin-dependent optical lattices \cite{meng2023} and techniques for observing vortex structures \cite{sadler2006}, our findings can be readily tested using current cold atom platforms and are expected to inspire further research on simulating twisted trilayer lattices in spin-1 BEC.

We thank Zuo Wang for helpful discussions. 
This work is supported by the National Key Research and Development Program of China (Grant No. 2022YFA1405304) and the National Natural Science Foundation of China (Grant No. 12004118). 

\appendix
\begin{figure*}[tb] \centering 
\centering
\includegraphics[width=0.99\textwidth]{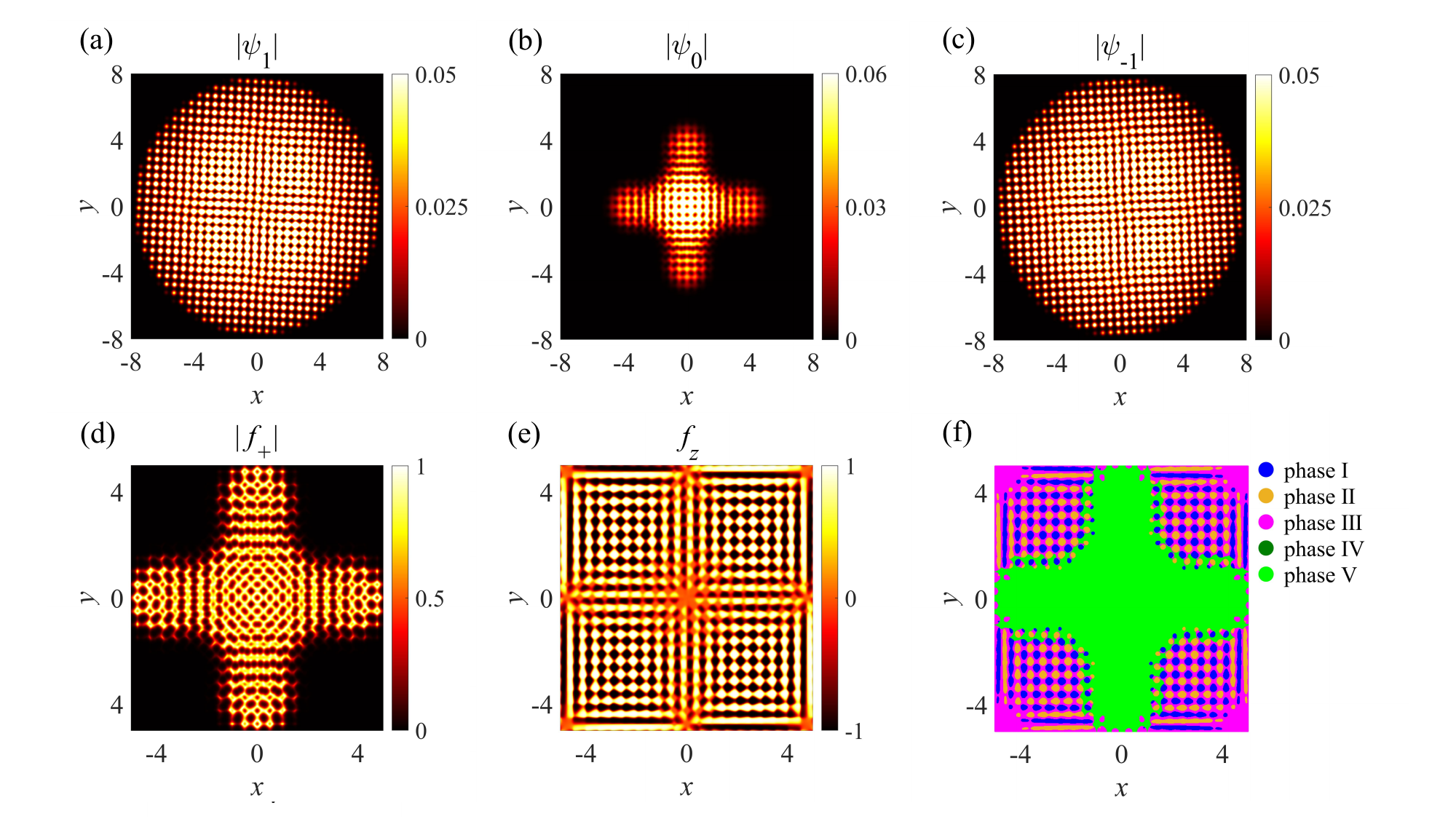}
	\caption{(Color online) The ground state properties of ferromagnetic spin-1 BEC in twisted optical lattice potentials. 
		(a)-(c) show the ground state density distribution of spin-1 BEC corresponding to the
		$m=1$, $m=0$ and $m=-1$ component, respectively. 
		The spatial distribution of transverse magnetization $\left| f_+\right|$ and longitudinal magnetization $f_z$ are shown in (d) and (e), respectively. (f) shows the spatial distribution of the ground state phases for $\epsilon=\delta=0.05$. The interaction strength $C_1 n=-0.007E_r$ and other parameters are the same as those in Fig. \ref{fig3}.}
	\label{fig6}
\end{figure*}

\section{Ground state properties and topological excitations of ferromagnetic spin-1 BEC in twisted optical lattices}
\label{Sec6}
In the main body of this paper, we discussed the ground state characteristics and quench dynamics under antiferromagnetic interaction in detail. In this appendix, we focus on the case of ferromagnetic interaction $(C_1<0)$. We find that the main features of the ground state in the ferromagnetic interaction case are similar to those observed in the antiferromagnetic case. Specifically, Figs. \ref{fig6}(a)-(e) illustrate the density and magnetization distribution under ferromagnetic interaction. Notably, all these physical quantities exhibit moir\'e periodicity.

\begin{figure*}[tb] \centering
	\includegraphics[width=0.99\linewidth]{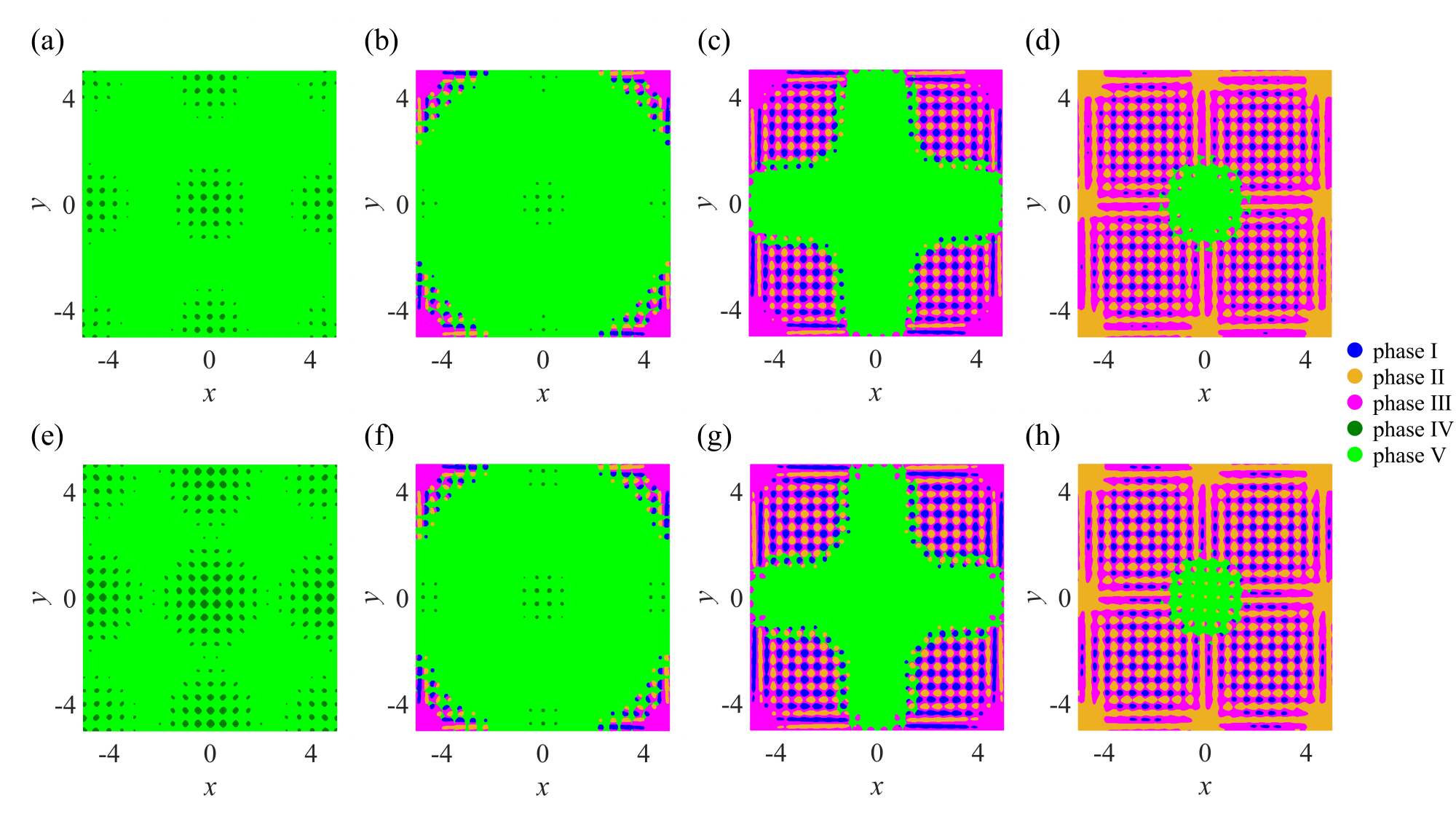}
	\caption{(Color online) The ground state phases of ferromagnetic spin-1 BEC under different lattice depths. The first and second rows correspond to threshold $\epsilon=\delta=0.03$ and $\epsilon=\delta=0.05$, respectively. The interaction strength $C_1 n=-0.007E_r$ and other parameters are the same as those in Fig. \ref{fig4}.}
	\label{fig7}
\end{figure*}

We use similar method to characterize the local phases of the ground state in the ferromagnetic case. For $\epsilon=\delta=0.05$, the spatial distribution of the ground state phases resembles the antiferromagnetic interaction case as shown in Fig. \ref{fig6}(f). At some specific positions in space, atoms are distributed only in $m=1$ and $m=-1$ components. This indicates the presence of the AFM phase which is not present in a homogeneous system under ferromagnetic interaction. The additional presence of AFM phase here can be explained by similar mechanism as the antiferromagnetic case, due to the spatial inhomogeneity induced kinetic energy. We also study the local phase distribution of the system under different lattice depths and thresholds for the ferromagnetic interaction, as shown in Figs. \ref{fig7}(a)-(h). By changing the lattice depth, all phases can also coexist under ferromagnetic interaction.

\begin{figure*}[tb] \centering
	\centering
	\includegraphics[width=0.99\linewidth]{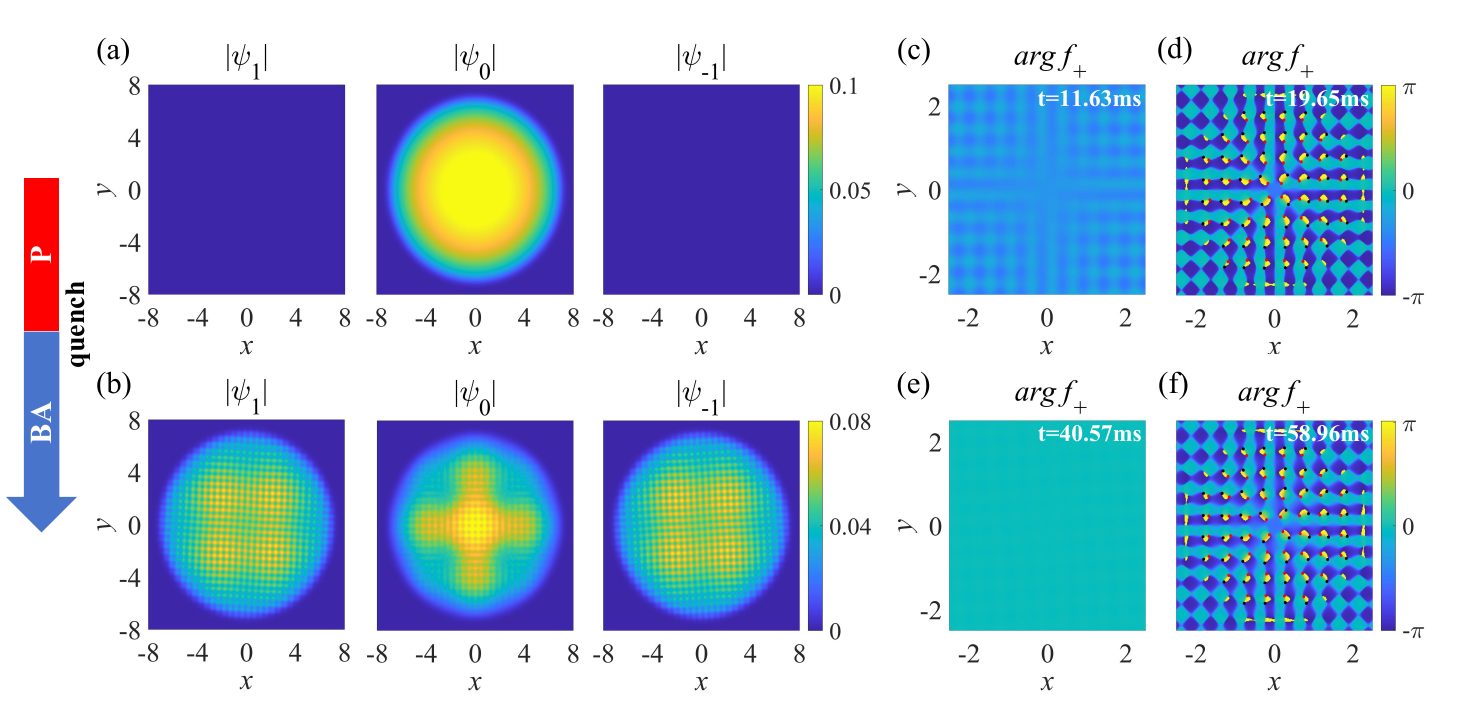}
	\caption{(Color online) Evolution of wave functions and magnetization of ferromagnetic spin-1 BEC in the twisted optical lattice following the instantaneous quench from $V_0=1.5E_r$ to $V_0=1.15E_r$ (shown by the arrow). (a), (b) The evolution of density distribution for P phase and BA phase, respectively. (c)-(f) The time evolution of the angle of $f_+$ ($arg f_+$) at different moments. The interaction strength $C_1n=-0.0007E_r$ and other parameters are the same as those in Fig. \ref{fig5}.}
	\label{fig8}
\end{figure*}

Finally, we study the topological excitations of the system under ferromagnetic interaction as shown in Fig. \ref{fig8}. The ground state of the system is prepared in the P phase and the transition from the P phase to the BA phase is realized by sudden quenching the lattice depth. The results show that the system can also excite periodic oscillation of vortex pairs under ferromagnetic interaction.

\bibliographystyle{apsrev4-2}

%apsrev4-2.bst 2019-01-14 (MD) hand-edited version of apsrev4-1.bst
%Control: key (0)
%Control: author (72) initials jnrlst
%Control: editor formatted (1) identically to author
%Control: production of article title (-1) disabled
%Control: page (0) single
%Control: year (1) truncated
%Control: production of eprint (0) enabled
\begin{thebibliography}{55}%
\makeatletter
\providecommand \@ifxundefined [1]{%
 \@ifx{#1\undefined}
}%
\providecommand \@ifnum [1]{%
 \ifnum #1\expandafter \@firstoftwo
 \else \expandafter \@secondoftwo
 \fi
}%
\providecommand \@ifx [1]{%
 \ifx #1\expandafter \@firstoftwo
 \else \expandafter \@secondoftwo
 \fi
}%
\providecommand \natexlab [1]{#1}%
\providecommand \enquote  [1]{``#1''}%
\providecommand \bibnamefont  [1]{#1}%
\providecommand \bibfnamefont [1]{#1}%
\providecommand \citenamefont [1]{#1}%
\providecommand \href@noop [0]{\@secondoftwo}%
\providecommand \href [0]{\begingroup \@sanitize@url \@href}%
\providecommand \@href[1]{\@@startlink{#1}\@@href}%
\providecommand \@@href[1]{\endgroup#1\@@endlink}%
\providecommand \@sanitize@url [0]{\catcode `\\12\catcode `\$12\catcode `\&12\catcode `\#12\catcode `\^12\catcode `\_12\catcode `\%12\relax}%
\providecommand \@@startlink[1]{}%
\providecommand \@@endlink[0]{}%
\providecommand \url  [0]{\begingroup\@sanitize@url \@url }%
\providecommand \@url [1]{\endgroup\@href {#1}{\urlprefix }}%
\providecommand \urlprefix  [0]{URL }%
\providecommand \Eprint [0]{\href }%
\providecommand \doibase [0]{https://doi.org/}%
\providecommand \selectlanguage [0]{\@gobble}%
\providecommand \bibinfo  [0]{\@secondoftwo}%
\providecommand \bibfield  [0]{\@secondoftwo}%
\providecommand \translation [1]{[#1]}%
\providecommand \BibitemOpen [0]{}%
\providecommand \bibitemStop [0]{}%
\providecommand \bibitemNoStop [0]{.\EOS\space}%
\providecommand \EOS [0]{\spacefactor3000\relax}%
\providecommand \BibitemShut  [1]{\csname bibitem#1\endcsname}%
\let\auto@bib@innerbib\@empty
%</preamble>
\bibitem [{\citenamefont {Cao}\ \emph {et~al.}(2018{\natexlab{a}})\citenamefont {Cao}, \citenamefont {Fatemi}, \citenamefont {Fang}, \citenamefont {Watanabe}, \citenamefont {Taniguchi}, \citenamefont {Kaxiras},\ and\ \citenamefont {{Jarillo-Herrero}}}]{cao2018}%
  \BibitemOpen
  \bibfield  {author} {\bibinfo {author} {\bibfnamefont {Y.}~\bibnamefont {Cao}}, \bibinfo {author} {\bibfnamefont {V.}~\bibnamefont {Fatemi}}, \bibinfo {author} {\bibfnamefont {S.}~\bibnamefont {Fang}}, \bibinfo {author} {\bibfnamefont {K.}~\bibnamefont {Watanabe}}, \bibinfo {author} {\bibfnamefont {T.}~\bibnamefont {Taniguchi}}, \bibinfo {author} {\bibfnamefont {E.}~\bibnamefont {Kaxiras}},\ and\ \bibinfo {author} {\bibfnamefont {P.}~\bibnamefont {{Jarillo-Herrero}}},\ }\href@noop {} {\bibfield  {journal} {\bibinfo  {journal} {Nature}\ }\textbf {\bibinfo {volume} {556}},\ \bibinfo {pages} {43} (\bibinfo {year} {2018}{\natexlab{a}})}\BibitemShut {NoStop}%
\bibitem [{\citenamefont {Cao}\ \emph {et~al.}(2018{\natexlab{b}})\citenamefont {Cao}, \citenamefont {Fatemi}, \citenamefont {Demir}, \citenamefont {Fang}, \citenamefont {Tomarken}, \citenamefont {Luo}, \citenamefont {{Sanchez-Yamagishi}}, \citenamefont {Watanabe}, \citenamefont {Taniguchi}, \citenamefont {Kaxiras}, \citenamefont {Ashoori},\ and\ \citenamefont {{Jarillo-Herrero}}}]{caoCorrelatedInsulatorBehaviour2018a}%
  \BibitemOpen
  \bibfield  {author} {\bibinfo {author} {\bibfnamefont {Y.}~\bibnamefont {Cao}}, \bibinfo {author} {\bibfnamefont {V.}~\bibnamefont {Fatemi}}, \bibinfo {author} {\bibfnamefont {A.}~\bibnamefont {Demir}}, \bibinfo {author} {\bibfnamefont {S.}~\bibnamefont {Fang}}, \bibinfo {author} {\bibfnamefont {S.~L.}\ \bibnamefont {Tomarken}}, \bibinfo {author} {\bibfnamefont {J.~Y.}\ \bibnamefont {Luo}}, \bibinfo {author} {\bibfnamefont {J.~D.}\ \bibnamefont {{Sanchez-Yamagishi}}}, \bibinfo {author} {\bibfnamefont {K.}~\bibnamefont {Watanabe}}, \bibinfo {author} {\bibfnamefont {T.}~\bibnamefont {Taniguchi}}, \bibinfo {author} {\bibfnamefont {E.}~\bibnamefont {Kaxiras}}, \bibinfo {author} {\bibfnamefont {R.~C.}\ \bibnamefont {Ashoori}},\ and\ \bibinfo {author} {\bibfnamefont {P.}~\bibnamefont {{Jarillo-Herrero}}},\ }\href@noop {} {\bibfield  {journal} {\bibinfo  {journal} {Nature}\ }\textbf {\bibinfo {volume} {556}},\ \bibinfo {pages} {80} (\bibinfo {year} {2018}{\natexlab{b}})}\BibitemShut {NoStop}%
\bibitem [{\citenamefont {Bistritzer}\ and\ \citenamefont {MacDonald}(2011)}]{bistritzer2011}%
  \BibitemOpen
  \bibfield  {author} {\bibinfo {author} {\bibfnamefont {R.}~\bibnamefont {Bistritzer}}\ and\ \bibinfo {author} {\bibfnamefont {A.~H.}\ \bibnamefont {MacDonald}},\ }\href@noop {} {\bibfield  {journal} {\bibinfo  {journal} {Proc. Natl. Acad. Sci. U.S.A.}\ }\textbf {\bibinfo {volume} {108}},\ \bibinfo {pages} {12233} (\bibinfo {year} {2011})}\BibitemShut {NoStop}%
\bibitem [{\citenamefont {Chen}\ \emph {et~al.}(2019)\citenamefont {Chen}, \citenamefont {Jiang}, \citenamefont {Wu}, \citenamefont {Lyu}, \citenamefont {Li}, \citenamefont {Chittari}, \citenamefont {Watanabe}, \citenamefont {Taniguchi}, \citenamefont {Shi}, \citenamefont {Jung}, \citenamefont {Zhang},\ and\ \citenamefont {Wang}}]{chen2019}%
  \BibitemOpen
  \bibfield  {author} {\bibinfo {author} {\bibfnamefont {G.}~\bibnamefont {Chen}}, \bibinfo {author} {\bibfnamefont {L.}~\bibnamefont {Jiang}}, \bibinfo {author} {\bibfnamefont {S.}~\bibnamefont {Wu}}, \bibinfo {author} {\bibfnamefont {B.}~\bibnamefont {Lyu}}, \bibinfo {author} {\bibfnamefont {H.}~\bibnamefont {Li}}, \bibinfo {author} {\bibfnamefont {B.~L.}\ \bibnamefont {Chittari}}, \bibinfo {author} {\bibfnamefont {K.}~\bibnamefont {Watanabe}}, \bibinfo {author} {\bibfnamefont {T.}~\bibnamefont {Taniguchi}}, \bibinfo {author} {\bibfnamefont {Z.}~\bibnamefont {Shi}}, \bibinfo {author} {\bibfnamefont {J.}~\bibnamefont {Jung}}, \bibinfo {author} {\bibfnamefont {Y.}~\bibnamefont {Zhang}},\ and\ \bibinfo {author} {\bibfnamefont {F.}~\bibnamefont {Wang}},\ }\href@noop {} {\bibfield  {journal} {\bibinfo  {journal} {Nat. Phys.}\ }\textbf {\bibinfo {volume} {15}},\ \bibinfo {pages} {237} (\bibinfo {year} {2019})}\BibitemShut {NoStop}%
\bibitem [{\citenamefont {Codecido}\ \emph {et~al.}(2019)\citenamefont {Codecido}, \citenamefont {Wang}, \citenamefont {Koester}, \citenamefont {Che}, \citenamefont {Tian}, \citenamefont {Lv}, \citenamefont {Tran}, \citenamefont {Watanabe}, \citenamefont {Taniguchi}, \citenamefont {Zhang}, \citenamefont {Bockrath},\ and\ \citenamefont {Lau}}]{codecido2019}%
  \BibitemOpen
  \bibfield  {author} {\bibinfo {author} {\bibfnamefont {E.}~\bibnamefont {Codecido}}, \bibinfo {author} {\bibfnamefont {Q.}~\bibnamefont {Wang}}, \bibinfo {author} {\bibfnamefont {R.}~\bibnamefont {Koester}}, \bibinfo {author} {\bibfnamefont {S.}~\bibnamefont {Che}}, \bibinfo {author} {\bibfnamefont {H.}~\bibnamefont {Tian}}, \bibinfo {author} {\bibfnamefont {R.}~\bibnamefont {Lv}}, \bibinfo {author} {\bibfnamefont {S.}~\bibnamefont {Tran}}, \bibinfo {author} {\bibfnamefont {K.}~\bibnamefont {Watanabe}}, \bibinfo {author} {\bibfnamefont {T.}~\bibnamefont {Taniguchi}}, \bibinfo {author} {\bibfnamefont {F.}~\bibnamefont {Zhang}}, \bibinfo {author} {\bibfnamefont {M.}~\bibnamefont {Bockrath}},\ and\ \bibinfo {author} {\bibfnamefont {C.~N.}\ \bibnamefont {Lau}},\ }\href@noop {} {\bibfield  {journal} {\bibinfo  {journal} {Sci. Adv.}\ }\textbf {\bibinfo {volume} {5}},\ \bibinfo {pages} {eaaw9770} (\bibinfo {year} {2019})}\BibitemShut {NoStop}%
\bibitem [{\citenamefont {Sharpe}\ \emph {et~al.}(2019)\citenamefont {Sharpe}, \citenamefont {Fox}, \citenamefont {Barnard}, \citenamefont {Finney}, \citenamefont {Watanabe}, \citenamefont {Taniguchi}, \citenamefont {Kastner},\ and\ \citenamefont {{Goldhaber-Gordon}}}]{sharpe2019}%
  \BibitemOpen
  \bibfield  {author} {\bibinfo {author} {\bibfnamefont {A.~L.}\ \bibnamefont {Sharpe}}, \bibinfo {author} {\bibfnamefont {E.~J.}\ \bibnamefont {Fox}}, \bibinfo {author} {\bibfnamefont {A.~W.}\ \bibnamefont {Barnard}}, \bibinfo {author} {\bibfnamefont {J.}~\bibnamefont {Finney}}, \bibinfo {author} {\bibfnamefont {K.}~\bibnamefont {Watanabe}}, \bibinfo {author} {\bibfnamefont {T.}~\bibnamefont {Taniguchi}}, \bibinfo {author} {\bibfnamefont {M.~A.}\ \bibnamefont {Kastner}},\ and\ \bibinfo {author} {\bibfnamefont {D.}~\bibnamefont {{Goldhaber-Gordon}}},\ }\href@noop {} {\bibfield  {journal} {\bibinfo  {journal} {Science}\ }\textbf {\bibinfo {volume} {365}},\ \bibinfo {pages} {605} (\bibinfo {year} {2019})}\BibitemShut {NoStop}%
\bibitem [{\citenamefont {Kerelsky}\ \emph {et~al.}(2019)\citenamefont {Kerelsky}, \citenamefont {McGilly}, \citenamefont {Kennes}, \citenamefont {Xian}, \citenamefont {Yankowitz}, \citenamefont {Chen}, \citenamefont {Watanabe}, \citenamefont {Taniguchi}, \citenamefont {Hone}, \citenamefont {Dean}, \citenamefont {Rubio},\ and\ \citenamefont {Pasupathy}}]{kerelsky2019}%
  \BibitemOpen
  \bibfield  {author} {\bibinfo {author} {\bibfnamefont {A.}~\bibnamefont {Kerelsky}}, \bibinfo {author} {\bibfnamefont {L.~J.}\ \bibnamefont {McGilly}}, \bibinfo {author} {\bibfnamefont {D.~M.}\ \bibnamefont {Kennes}}, \bibinfo {author} {\bibfnamefont {L.}~\bibnamefont {Xian}}, \bibinfo {author} {\bibfnamefont {M.}~\bibnamefont {Yankowitz}}, \bibinfo {author} {\bibfnamefont {S.}~\bibnamefont {Chen}}, \bibinfo {author} {\bibfnamefont {K.}~\bibnamefont {Watanabe}}, \bibinfo {author} {\bibfnamefont {T.}~\bibnamefont {Taniguchi}}, \bibinfo {author} {\bibfnamefont {J.}~\bibnamefont {Hone}}, \bibinfo {author} {\bibfnamefont {C.}~\bibnamefont {Dean}}, \bibinfo {author} {\bibfnamefont {A.}~\bibnamefont {Rubio}},\ and\ \bibinfo {author} {\bibfnamefont {A.~N.}\ \bibnamefont {Pasupathy}},\ }\href@noop {} {\bibfield  {journal} {\bibinfo  {journal} {Nature}\ }\textbf {\bibinfo {volume} {572}},\ \bibinfo {pages} {95} (\bibinfo {year} {2019})}\BibitemShut {NoStop}%
\bibitem [{\citenamefont {Wilson}\ \emph {et~al.}(2021)\citenamefont {Wilson}, \citenamefont {Yao}, \citenamefont {Shan},\ and\ \citenamefont {Xu}}]{wilson2021}%
  \BibitemOpen
  \bibfield  {author} {\bibinfo {author} {\bibfnamefont {N.~P.}\ \bibnamefont {Wilson}}, \bibinfo {author} {\bibfnamefont {W.}~\bibnamefont {Yao}}, \bibinfo {author} {\bibfnamefont {J.}~\bibnamefont {Shan}},\ and\ \bibinfo {author} {\bibfnamefont {X.}~\bibnamefont {Xu}},\ }\href@noop {} {\bibfield  {journal} {\bibinfo  {journal} {Nature}\ }\textbf {\bibinfo {volume} {599}},\ \bibinfo {pages} {383} (\bibinfo {year} {2021})}\BibitemShut {NoStop}%
\bibitem [{\citenamefont {Li}\ \emph {et~al.}(2021)\citenamefont {Li}, \citenamefont {Jiang}, \citenamefont {Shen}, \citenamefont {Zhang}, \citenamefont {Li}, \citenamefont {Tao}, \citenamefont {Devakul}, \citenamefont {Watanabe}, \citenamefont {Taniguchi}, \citenamefont {Fu}, \citenamefont {Shan},\ and\ \citenamefont {Mak}}]{li2021}%
  \BibitemOpen
  \bibfield  {author} {\bibinfo {author} {\bibfnamefont {T.}~\bibnamefont {Li}}, \bibinfo {author} {\bibfnamefont {S.}~\bibnamefont {Jiang}}, \bibinfo {author} {\bibfnamefont {B.}~\bibnamefont {Shen}}, \bibinfo {author} {\bibfnamefont {Y.}~\bibnamefont {Zhang}}, \bibinfo {author} {\bibfnamefont {L.}~\bibnamefont {Li}}, \bibinfo {author} {\bibfnamefont {Z.}~\bibnamefont {Tao}}, \bibinfo {author} {\bibfnamefont {T.}~\bibnamefont {Devakul}}, \bibinfo {author} {\bibfnamefont {K.}~\bibnamefont {Watanabe}}, \bibinfo {author} {\bibfnamefont {T.}~\bibnamefont {Taniguchi}}, \bibinfo {author} {\bibfnamefont {L.}~\bibnamefont {Fu}}, \bibinfo {author} {\bibfnamefont {J.}~\bibnamefont {Shan}},\ and\ \bibinfo {author} {\bibfnamefont {K.~F.}\ \bibnamefont {Mak}},\ }\href@noop {} {\bibfield  {journal} {\bibinfo  {journal} {Nature}\ }\textbf {\bibinfo {volume} {600}},\ \bibinfo {pages} {641} (\bibinfo {year} {2021})}\BibitemShut {NoStop}%
\bibitem [{\citenamefont {Yu}\ \emph {et~al.}(2019)\citenamefont {Yu}, \citenamefont {Chen},\ and\ \citenamefont {Yao}}]{yu2019giant}%
  \BibitemOpen
  \bibfield  {author} {\bibinfo {author} {\bibfnamefont {H.}~\bibnamefont {Yu}}, \bibinfo {author} {\bibfnamefont {M.}~\bibnamefont {Chen}},\ and\ \bibinfo {author} {\bibfnamefont {W.}~\bibnamefont {Yao}},\ }\href@noop {} {\bibfield  {journal} {\bibinfo  {journal} {Natl. Sci. Rev.}\ }\textbf {\bibinfo {volume} {7}},\ \bibinfo {pages} {12} (\bibinfo {year} {2019})}\BibitemShut {NoStop}%
\bibitem [{\citenamefont {Wu}\ \emph {et~al.}(2019)\citenamefont {Wu}, \citenamefont {Lovorn}, \citenamefont {Tutuc}, \citenamefont {Martin},\ and\ \citenamefont {MacDonald}}]{wu2019}%
  \BibitemOpen
  \bibfield  {author} {\bibinfo {author} {\bibfnamefont {F.}~\bibnamefont {Wu}}, \bibinfo {author} {\bibfnamefont {T.}~\bibnamefont {Lovorn}}, \bibinfo {author} {\bibfnamefont {E.}~\bibnamefont {Tutuc}}, \bibinfo {author} {\bibfnamefont {I.}~\bibnamefont {Martin}},\ and\ \bibinfo {author} {\bibfnamefont {A.~H.}\ \bibnamefont {MacDonald}},\ }\href@noop {} {\bibfield  {journal} {\bibinfo  {journal} {Phys. Rev. Lett.}\ }\textbf {\bibinfo {volume} {122}},\ \bibinfo {pages} {086402} (\bibinfo {year} {2019})}\BibitemShut {NoStop}%
\bibitem [{\citenamefont {Devakul}\ \emph {et~al.}(2021)\citenamefont {Devakul}, \citenamefont {Cr{\'e}pel}, \citenamefont {Zhang},\ and\ \citenamefont {Fu}}]{devakul2021}%
  \BibitemOpen
  \bibfield  {author} {\bibinfo {author} {\bibfnamefont {T.}~\bibnamefont {Devakul}}, \bibinfo {author} {\bibfnamefont {V.}~\bibnamefont {Cr{\'e}pel}}, \bibinfo {author} {\bibfnamefont {Y.}~\bibnamefont {Zhang}},\ and\ \bibinfo {author} {\bibfnamefont {L.}~\bibnamefont {Fu}},\ }\href@noop {} {\bibfield  {journal} {\bibinfo  {journal} {Nat. Commun.}\ }\textbf {\bibinfo {volume} {12}},\ \bibinfo {pages} {6730} (\bibinfo {year} {2021})}\BibitemShut {NoStop}%
\bibitem [{\citenamefont {Zeng}\ \emph {et~al.}(2023)\citenamefont {Zeng}, \citenamefont {Xia}, \citenamefont {Kang}, \citenamefont {Zhu}, \citenamefont {Kn{\"u}ppel}, \citenamefont {Vaswani}, \citenamefont {Watanabe}, \citenamefont {Taniguchi}, \citenamefont {Mak},\ and\ \citenamefont {Shan}}]{zeng2023}%
  \BibitemOpen
  \bibfield  {author} {\bibinfo {author} {\bibfnamefont {Y.}~\bibnamefont {Zeng}}, \bibinfo {author} {\bibfnamefont {Z.}~\bibnamefont {Xia}}, \bibinfo {author} {\bibfnamefont {K.}~\bibnamefont {Kang}}, \bibinfo {author} {\bibfnamefont {J.}~\bibnamefont {Zhu}}, \bibinfo {author} {\bibfnamefont {P.}~\bibnamefont {Kn{\"u}ppel}}, \bibinfo {author} {\bibfnamefont {C.}~\bibnamefont {Vaswani}}, \bibinfo {author} {\bibfnamefont {K.}~\bibnamefont {Watanabe}}, \bibinfo {author} {\bibfnamefont {T.}~\bibnamefont {Taniguchi}}, \bibinfo {author} {\bibfnamefont {K.~F.}\ \bibnamefont {Mak}},\ and\ \bibinfo {author} {\bibfnamefont {J.}~\bibnamefont {Shan}},\ }\href@noop {} {\bibfield  {journal} {\bibinfo  {journal} {Nature}\ }\textbf {\bibinfo {volume} {622}},\ \bibinfo {pages} {69} (\bibinfo {year} {2023})}\BibitemShut {NoStop}%
\bibitem [{\citenamefont {Xu}\ \emph {et~al.}(2023)\citenamefont {Xu}, \citenamefont {Sun}, \citenamefont {Jia}, \citenamefont {Liu}, \citenamefont {Xu}, \citenamefont {Li}, \citenamefont {Gu}, \citenamefont {Watanabe}, \citenamefont {Taniguchi}, \citenamefont {Tong}, \citenamefont {Jia}, \citenamefont {Shi}, \citenamefont {Jiang}, \citenamefont {Zhang}, \citenamefont {Liu},\ and\ \citenamefont {Li}}]{xu2023}%
  \BibitemOpen
  \bibfield  {author} {\bibinfo {author} {\bibfnamefont {F.}~\bibnamefont {Xu}}, \bibinfo {author} {\bibfnamefont {Z.}~\bibnamefont {Sun}}, \bibinfo {author} {\bibfnamefont {T.}~\bibnamefont {Jia}}, \bibinfo {author} {\bibfnamefont {C.}~\bibnamefont {Liu}}, \bibinfo {author} {\bibfnamefont {C.}~\bibnamefont {Xu}}, \bibinfo {author} {\bibfnamefont {C.}~\bibnamefont {Li}}, \bibinfo {author} {\bibfnamefont {Y.}~\bibnamefont {Gu}}, \bibinfo {author} {\bibfnamefont {K.}~\bibnamefont {Watanabe}}, \bibinfo {author} {\bibfnamefont {T.}~\bibnamefont {Taniguchi}}, \bibinfo {author} {\bibfnamefont {B.}~\bibnamefont {Tong}}, \bibinfo {author} {\bibfnamefont {J.}~\bibnamefont {Jia}}, \bibinfo {author} {\bibfnamefont {Z.}~\bibnamefont {Shi}}, \bibinfo {author} {\bibfnamefont {S.}~\bibnamefont {Jiang}}, \bibinfo {author} {\bibfnamefont {Y.}~\bibnamefont {Zhang}}, \bibinfo {author} {\bibfnamefont {X.}~\bibnamefont {Liu}},\ and\ \bibinfo {author} {\bibfnamefont {T.}~\bibnamefont {Li}},\ }\href@noop {} {\bibfield  {journal} {\bibinfo  {journal} {Phys. Rev. X}\ }\textbf {\bibinfo {volume} {13}},\ \bibinfo {pages} {031037} (\bibinfo {year} {2023})}\BibitemShut {NoStop}%
\bibitem [{\citenamefont {Li}\ \emph {et~al.}(2024)\citenamefont {Li}, \citenamefont {Qiu},\ and\ \citenamefont {Wu}}]{li2024}%
  \BibitemOpen
  \bibfield  {author} {\bibinfo {author} {\bibfnamefont {B.}~\bibnamefont {Li}}, \bibinfo {author} {\bibfnamefont {W.-X.}\ \bibnamefont {Qiu}},\ and\ \bibinfo {author} {\bibfnamefont {F.}~\bibnamefont {Wu}},\ }\href@noop {} {\bibfield  {journal} {\bibinfo  {journal} {Phys. Rev. B}\ }\textbf {\bibinfo {volume} {109}},\ \bibinfo {pages} {L041106} (\bibinfo {year} {2024})}\BibitemShut {NoStop}%
\bibitem [{\citenamefont {Park}\ \emph {et~al.}(2023)\citenamefont {Park}, \citenamefont {Cai}, \citenamefont {Anderson}, \citenamefont {Zhang}, \citenamefont {Zhu}, \citenamefont {Liu}, \citenamefont {Wang}, \citenamefont {Holtzmann}, \citenamefont {Hu}, \citenamefont {Liu}, \citenamefont {Taniguchi}, \citenamefont {Watanabe}, \citenamefont {Chu}, \citenamefont {Cao}, \citenamefont {Fu}, \citenamefont {Yao}, \citenamefont {Chang}, \citenamefont {Cobden}, \citenamefont {Xiao},\ and\ \citenamefont {Xu}}]{Liu2023a}%
  \BibitemOpen
  \bibfield  {author} {\bibinfo {author} {\bibfnamefont {H.}~\bibnamefont {Park}}, \bibinfo {author} {\bibfnamefont {J.}~\bibnamefont {Cai}}, \bibinfo {author} {\bibfnamefont {E.}~\bibnamefont {Anderson}}, \bibinfo {author} {\bibfnamefont {Y.}~\bibnamefont {Zhang}}, \bibinfo {author} {\bibfnamefont {J.}~\bibnamefont {Zhu}}, \bibinfo {author} {\bibfnamefont {X.}~\bibnamefont {Liu}}, \bibinfo {author} {\bibfnamefont {C.}~\bibnamefont {Wang}}, \bibinfo {author} {\bibfnamefont {W.}~\bibnamefont {Holtzmann}}, \bibinfo {author} {\bibfnamefont {C.}~\bibnamefont {Hu}}, \bibinfo {author} {\bibfnamefont {Z.}~\bibnamefont {Liu}}, \bibinfo {author} {\bibfnamefont {T.}~\bibnamefont {Taniguchi}}, \bibinfo {author} {\bibfnamefont {K.}~\bibnamefont {Watanabe}}, \bibinfo {author} {\bibfnamefont {J.-H.}\ \bibnamefont {Chu}}, \bibinfo {author} {\bibfnamefont {T.}~\bibnamefont {Cao}}, \bibinfo {author} {\bibfnamefont {L.}~\bibnamefont {Fu}}, \bibinfo {author} {\bibfnamefont {W.}~\bibnamefont {Yao}}, \bibinfo {author} {\bibfnamefont {C.-Z.}\ \bibnamefont {Chang}}, \bibinfo {author} {\bibfnamefont {D.}~\bibnamefont {Cobden}}, \bibinfo {author} {\bibfnamefont {D.}~\bibnamefont {Xiao}},\ and\ \bibinfo {author} {\bibfnamefont {X.}~\bibnamefont {Xu}},\ }\href@noop {} {\bibfield  {journal} {\bibinfo  {journal} {Nature}\ }\textbf {\bibinfo {volume} {622}},\ \bibinfo {pages} {74} (\bibinfo {year} {2023})}\BibitemShut {NoStop}%
\bibitem [{\citenamefont {Cai}\ \emph {et~al.}(2023)\citenamefont {Cai}, \citenamefont {Anderson}, \citenamefont {Wang}, \citenamefont {Zhang}, \citenamefont {Liu}, \citenamefont {Holtzmann}, \citenamefont {Zhang}, \citenamefont {Fan}, \citenamefont {Taniguchi}, \citenamefont {Watanabe}, \citenamefont {Ran}, \citenamefont {Cao}, \citenamefont {Fu}, \citenamefont {Xiao}, \citenamefont {Yao},\ and\ \citenamefont {Xu}}]{cai2023nature}%
  \BibitemOpen
  \bibfield  {author} {\bibinfo {author} {\bibfnamefont {J.}~\bibnamefont {Cai}}, \bibinfo {author} {\bibfnamefont {E.}~\bibnamefont {Anderson}}, \bibinfo {author} {\bibfnamefont {C.}~\bibnamefont {Wang}}, \bibinfo {author} {\bibfnamefont {X.}~\bibnamefont {Zhang}}, \bibinfo {author} {\bibfnamefont {X.}~\bibnamefont {Liu}}, \bibinfo {author} {\bibfnamefont {W.}~\bibnamefont {Holtzmann}}, \bibinfo {author} {\bibfnamefont {Y.}~\bibnamefont {Zhang}}, \bibinfo {author} {\bibfnamefont {F.}~\bibnamefont {Fan}}, \bibinfo {author} {\bibfnamefont {T.}~\bibnamefont {Taniguchi}}, \bibinfo {author} {\bibfnamefont {K.}~\bibnamefont {Watanabe}}, \bibinfo {author} {\bibfnamefont {Y.}~\bibnamefont {Ran}}, \bibinfo {author} {\bibfnamefont {T.}~\bibnamefont {Cao}}, \bibinfo {author} {\bibfnamefont {L.}~\bibnamefont {Fu}}, \bibinfo {author} {\bibfnamefont {D.}~\bibnamefont {Xiao}}, \bibinfo {author} {\bibfnamefont {W.}~\bibnamefont {Yao}},\ and\ \bibinfo {author} {\bibfnamefont {X.}~\bibnamefont {Xu}},\ }\href@noop {} {\bibfield  {journal} {\bibinfo  {journal} {Nature}\ }\textbf {\bibinfo {volume} {622}},\ \bibinfo {pages} {63} (\bibinfo {year} {2023})}\BibitemShut {NoStop}%
\bibitem [{\citenamefont {Huang}\ \emph {et~al.}(2016)\citenamefont {Huang}, \citenamefont {Ye}, \citenamefont {Chen}, \citenamefont {Kartashov}, \citenamefont {Konotop},\ and\ \citenamefont {Torner}}]{huang2016}%
  \BibitemOpen
  \bibfield  {author} {\bibinfo {author} {\bibfnamefont {C.}~\bibnamefont {Huang}}, \bibinfo {author} {\bibfnamefont {F.}~\bibnamefont {Ye}}, \bibinfo {author} {\bibfnamefont {X.}~\bibnamefont {Chen}}, \bibinfo {author} {\bibfnamefont {Y.~V.}\ \bibnamefont {Kartashov}}, \bibinfo {author} {\bibfnamefont {V.~V.}\ \bibnamefont {Konotop}},\ and\ \bibinfo {author} {\bibfnamefont {L.}~\bibnamefont {Torner}},\ }\href@noop {} {\bibfield  {journal} {\bibinfo  {journal} {Sci. Rep.}\ }\textbf {\bibinfo {volume} {6}},\ \bibinfo {pages} {32546} (\bibinfo {year} {2016})}\BibitemShut {NoStop}%
\bibitem [{\citenamefont {Wang}\ \emph {et~al.}(2020)\citenamefont {Wang}, \citenamefont {Zheng}, \citenamefont {Chen}, \citenamefont {Huang}, \citenamefont {Kartashov}, \citenamefont {Torner}, \citenamefont {Konotop},\ and\ \citenamefont {Ye}}]{wang2020}%
  \BibitemOpen
  \bibfield  {author} {\bibinfo {author} {\bibfnamefont {P.}~\bibnamefont {Wang}}, \bibinfo {author} {\bibfnamefont {Y.}~\bibnamefont {Zheng}}, \bibinfo {author} {\bibfnamefont {X.}~\bibnamefont {Chen}}, \bibinfo {author} {\bibfnamefont {C.}~\bibnamefont {Huang}}, \bibinfo {author} {\bibfnamefont {Y.~V.}\ \bibnamefont {Kartashov}}, \bibinfo {author} {\bibfnamefont {L.}~\bibnamefont {Torner}}, \bibinfo {author} {\bibfnamefont {V.~V.}\ \bibnamefont {Konotop}},\ and\ \bibinfo {author} {\bibfnamefont {F.}~\bibnamefont {Ye}},\ }\href@noop {} {\bibfield  {journal} {\bibinfo  {journal} {Nature}\ }\textbf {\bibinfo {volume} {577}},\ \bibinfo {pages} {42} (\bibinfo {year} {2020})}\BibitemShut {NoStop}%
\bibitem [{\citenamefont {Fu}\ \emph {et~al.}(2020)\citenamefont {Fu}, \citenamefont {Wang}, \citenamefont {Huang}, \citenamefont {Kartashov}, \citenamefont {Torner}, \citenamefont {Konotop},\ and\ \citenamefont {Ye}}]{fu2020}%
  \BibitemOpen
  \bibfield  {author} {\bibinfo {author} {\bibfnamefont {Q.}~\bibnamefont {Fu}}, \bibinfo {author} {\bibfnamefont {P.}~\bibnamefont {Wang}}, \bibinfo {author} {\bibfnamefont {C.}~\bibnamefont {Huang}}, \bibinfo {author} {\bibfnamefont {Y.~V.}\ \bibnamefont {Kartashov}}, \bibinfo {author} {\bibfnamefont {L.}~\bibnamefont {Torner}}, \bibinfo {author} {\bibfnamefont {V.~V.}\ \bibnamefont {Konotop}},\ and\ \bibinfo {author} {\bibfnamefont {F.}~\bibnamefont {Ye}},\ }\href@noop {} {\bibfield  {journal} {\bibinfo  {journal} {Nat. Photonics}\ }\textbf {\bibinfo {volume} {14}},\ \bibinfo {pages} {663} (\bibinfo {year} {2020})}\BibitemShut {NoStop}%
\bibitem [{\citenamefont {Arkhipova}\ \emph {et~al.}(2023)\citenamefont {Arkhipova}, \citenamefont {Kartashov}, \citenamefont {Ivanov}, \citenamefont {Zhuravitskii}, \citenamefont {Skryabin}, \citenamefont {Dyakonov}, \citenamefont {Kalinkin}, \citenamefont {Kulik}, \citenamefont {Kompanets}, \citenamefont {Chekalin}, \citenamefont {Ye}, \citenamefont {Konotop}, \citenamefont {Torner},\ and\ \citenamefont {Zadkov}}]{arkhipova2023}%
  \BibitemOpen
  \bibfield  {author} {\bibinfo {author} {\bibfnamefont {A.~A.}\ \bibnamefont {Arkhipova}}, \bibinfo {author} {\bibfnamefont {Y.~V.}\ \bibnamefont {Kartashov}}, \bibinfo {author} {\bibfnamefont {S.~K.}\ \bibnamefont {Ivanov}}, \bibinfo {author} {\bibfnamefont {S.~A.}\ \bibnamefont {Zhuravitskii}}, \bibinfo {author} {\bibfnamefont {N.~N.}\ \bibnamefont {Skryabin}}, \bibinfo {author} {\bibfnamefont {I.~V.}\ \bibnamefont {Dyakonov}}, \bibinfo {author} {\bibfnamefont {A.~A.}\ \bibnamefont {Kalinkin}}, \bibinfo {author} {\bibfnamefont {S.~P.}\ \bibnamefont {Kulik}}, \bibinfo {author} {\bibfnamefont {V.~O.}\ \bibnamefont {Kompanets}}, \bibinfo {author} {\bibfnamefont {S.~V.}\ \bibnamefont {Chekalin}}, \bibinfo {author} {\bibfnamefont {F.}~\bibnamefont {Ye}}, \bibinfo {author} {\bibfnamefont {V.~V.}\ \bibnamefont {Konotop}}, \bibinfo {author} {\bibfnamefont {L.}~\bibnamefont {Torner}},\ and\ \bibinfo {author} {\bibfnamefont {V.~N.}\ \bibnamefont {Zadkov}},\ }\href@noop {} {\bibfield  {journal} {\bibinfo  {journal} {Phys. Rev. Lett.}\ }\textbf {\bibinfo {volume} {130}},\ \bibinfo {pages} {083801} (\bibinfo {year} {2023})}\BibitemShut {NoStop}%
\bibitem [{\citenamefont {Wu}\ \emph {et~al.}(2022)\citenamefont {Wu}, \citenamefont {Lin}, \citenamefont {Jiang}, \citenamefont {Zhou}, \citenamefont {Hang}, \citenamefont {Hou},\ and\ \citenamefont {Jiang}}]{wu2022a}%
  \BibitemOpen
  \bibfield  {author} {\bibinfo {author} {\bibfnamefont {S.-Q.}\ \bibnamefont {Wu}}, \bibinfo {author} {\bibfnamefont {Z.-K.}\ \bibnamefont {Lin}}, \bibinfo {author} {\bibfnamefont {B.}~\bibnamefont {Jiang}}, \bibinfo {author} {\bibfnamefont {X.}~\bibnamefont {Zhou}}, \bibinfo {author} {\bibfnamefont {Z.~H.}\ \bibnamefont {Hang}}, \bibinfo {author} {\bibfnamefont {B.}~\bibnamefont {Hou}},\ and\ \bibinfo {author} {\bibfnamefont {J.-H.}\ \bibnamefont {Jiang}},\ }\href@noop {} {\bibfield  {journal} {\bibinfo  {journal} {Phys. Rev. Appl.}\ }\textbf {\bibinfo {volume} {17}},\ \bibinfo {pages} {034061} (\bibinfo {year} {2022})}\BibitemShut {NoStop}%
\bibitem [{\citenamefont {Duan}\ \emph {et~al.}(2023)\citenamefont {Duan}, \citenamefont {Zheng}, \citenamefont {Zhang}, \citenamefont {Jiang}, \citenamefont {Man}, \citenamefont {Yu},\ and\ \citenamefont {Xia}}]{duan2023}%
  \BibitemOpen
  \bibfield  {author} {\bibinfo {author} {\bibfnamefont {G.}~\bibnamefont {Duan}}, \bibinfo {author} {\bibfnamefont {S.}~\bibnamefont {Zheng}}, \bibinfo {author} {\bibfnamefont {J.}~\bibnamefont {Zhang}}, \bibinfo {author} {\bibfnamefont {Z.}~\bibnamefont {Jiang}}, \bibinfo {author} {\bibfnamefont {X.}~\bibnamefont {Man}}, \bibinfo {author} {\bibfnamefont {D.}~\bibnamefont {Yu}},\ and\ \bibinfo {author} {\bibfnamefont {B.}~\bibnamefont {Xia}},\ }\href@noop {} {\bibfield  {journal} {\bibinfo  {journal} {Appl. Phys. Lett.}\ }\textbf {\bibinfo {volume} {123}},\ \bibinfo {pages} {021702} (\bibinfo {year} {2023})}\BibitemShut {NoStop}%
\bibitem [{\citenamefont {Oudich}\ \emph {et~al.}(2024)\citenamefont {Oudich}, \citenamefont {Kong}, \citenamefont {Zhang}, \citenamefont {Qiu},\ and\ \citenamefont {Jing}}]{oudich2024}%
  \BibitemOpen
  \bibfield  {author} {\bibinfo {author} {\bibfnamefont {M.}~\bibnamefont {Oudich}}, \bibinfo {author} {\bibfnamefont {X.}~\bibnamefont {Kong}}, \bibinfo {author} {\bibfnamefont {T.}~\bibnamefont {Zhang}}, \bibinfo {author} {\bibfnamefont {C.}~\bibnamefont {Qiu}},\ and\ \bibinfo {author} {\bibfnamefont {Y.}~\bibnamefont {Jing}},\ }\href@noop {} {\bibfield  {journal} {\bibinfo  {journal} {Nat. Mater.}\ }\textbf {\bibinfo {volume} {23}},\ \bibinfo {pages} {1169} (\bibinfo {year} {2024})}\BibitemShut {NoStop}%
\bibitem [{\citenamefont {{Soltan-Panahi}}\ \emph {et~al.}(2011)\citenamefont {{Soltan-Panahi}}, \citenamefont {Struck}, \citenamefont {Hauke}, \citenamefont {Bick}, \citenamefont {Plenkers}, \citenamefont {Meineke}, \citenamefont {Becker}, \citenamefont {Windpassinger}, \citenamefont {Lewenstein},\ and\ \citenamefont {Sengstock}}]{soltan-panahi2011a}%
  \BibitemOpen
  \bibfield  {author} {\bibinfo {author} {\bibfnamefont {P.}~\bibnamefont {{Soltan-Panahi}}}, \bibinfo {author} {\bibfnamefont {J.}~\bibnamefont {Struck}}, \bibinfo {author} {\bibfnamefont {P.}~\bibnamefont {Hauke}}, \bibinfo {author} {\bibfnamefont {A.}~\bibnamefont {Bick}}, \bibinfo {author} {\bibfnamefont {W.}~\bibnamefont {Plenkers}}, \bibinfo {author} {\bibfnamefont {G.}~\bibnamefont {Meineke}}, \bibinfo {author} {\bibfnamefont {C.}~\bibnamefont {Becker}}, \bibinfo {author} {\bibfnamefont {P.}~\bibnamefont {Windpassinger}}, \bibinfo {author} {\bibfnamefont {M.}~\bibnamefont {Lewenstein}},\ and\ \bibinfo {author} {\bibfnamefont {K.}~\bibnamefont {Sengstock}},\ }\href@noop {} {\bibfield  {journal} {\bibinfo  {journal} {Nat. Phys.}\ }\textbf {\bibinfo {volume} {7}},\ \bibinfo {pages} {434} (\bibinfo {year} {2011})}\BibitemShut {NoStop}%
\bibitem [{\citenamefont {Jo}\ \emph {et~al.}(2012)\citenamefont {Jo}, \citenamefont {Guzman}, \citenamefont {Thomas}, \citenamefont {Hosur}, \citenamefont {Vishwanath},\ and\ \citenamefont {{Stamper-Kurn}}}]{jo2012}%
  \BibitemOpen
  \bibfield  {author} {\bibinfo {author} {\bibfnamefont {G.-B.}\ \bibnamefont {Jo}}, \bibinfo {author} {\bibfnamefont {J.}~\bibnamefont {Guzman}}, \bibinfo {author} {\bibfnamefont {C.~K.}\ \bibnamefont {Thomas}}, \bibinfo {author} {\bibfnamefont {P.}~\bibnamefont {Hosur}}, \bibinfo {author} {\bibfnamefont {A.}~\bibnamefont {Vishwanath}},\ and\ \bibinfo {author} {\bibfnamefont {D.~M.}\ \bibnamefont {{Stamper-Kurn}}},\ }\href@noop {} {\bibfield  {journal} {\bibinfo  {journal} {Phys. Rev. Lett.}\ }\textbf {\bibinfo {volume} {108}},\ \bibinfo {pages} {045305} (\bibinfo {year} {2012})}\BibitemShut {NoStop}%
\bibitem [{\citenamefont {Luo}\ and\ \citenamefont {Zhang}(2021)}]{luo2021}%
  \BibitemOpen
  \bibfield  {author} {\bibinfo {author} {\bibfnamefont {X.-W.}\ \bibnamefont {Luo}}\ and\ \bibinfo {author} {\bibfnamefont {C.}~\bibnamefont {Zhang}},\ }\href@noop {} {\bibfield  {journal} {\bibinfo  {journal} {Phys. Rev. Lett.}\ }\textbf {\bibinfo {volume} {126}},\ \bibinfo {pages} {103201} (\bibinfo {year} {2021})}\BibitemShut {NoStop}%
\bibitem [{\citenamefont {Lee}\ and\ \citenamefont {Pixley}(2022)}]{Lee2022}%
  \BibitemOpen
  \bibfield  {author} {\bibinfo {author} {\bibfnamefont {J.}~\bibnamefont {Lee}}\ and\ \bibinfo {author} {\bibfnamefont {J.~H.}\ \bibnamefont {Pixley}},\ }\href@noop {} {\bibfield  {journal} {\bibinfo  {journal} {SciPost Phys.}\ }\textbf {\bibinfo {volume} {13}},\ \bibinfo {pages} {033} (\bibinfo {year} {2022})}\BibitemShut {NoStop}%
\bibitem [{\citenamefont {Meng}\ \emph {et~al.}(2023)\citenamefont {Meng}, \citenamefont {Wang}, \citenamefont {Han}, \citenamefont {Liu}, \citenamefont {Wen}, \citenamefont {Gao}, \citenamefont {Wang}, \citenamefont {Chin},\ and\ \citenamefont {Zhang}}]{meng2023}%
  \BibitemOpen
  \bibfield  {author} {\bibinfo {author} {\bibfnamefont {Z.}~\bibnamefont {Meng}}, \bibinfo {author} {\bibfnamefont {L.}~\bibnamefont {Wang}}, \bibinfo {author} {\bibfnamefont {W.}~\bibnamefont {Han}}, \bibinfo {author} {\bibfnamefont {F.}~\bibnamefont {Liu}}, \bibinfo {author} {\bibfnamefont {K.}~\bibnamefont {Wen}}, \bibinfo {author} {\bibfnamefont {C.}~\bibnamefont {Gao}}, \bibinfo {author} {\bibfnamefont {P.}~\bibnamefont {Wang}}, \bibinfo {author} {\bibfnamefont {C.}~\bibnamefont {Chin}},\ and\ \bibinfo {author} {\bibfnamefont {J.}~\bibnamefont {Zhang}},\ }\href@noop {} {\bibfield  {journal} {\bibinfo  {journal} {Nature}\ }\textbf {\bibinfo {volume} {615}},\ \bibinfo {pages} {231} (\bibinfo {year} {2023})}\BibitemShut {NoStop}%
\bibitem [{\citenamefont {{Gonz{\'a}lez-Tudela}}\ and\ \citenamefont {Cirac}(2019)}]{gonzalez-tudela2019}%
  \BibitemOpen
  \bibfield  {author} {\bibinfo {author} {\bibfnamefont {A.}~\bibnamefont {{Gonz{\'a}lez-Tudela}}}\ and\ \bibinfo {author} {\bibfnamefont {J.~I.}\ \bibnamefont {Cirac}},\ }\href@noop {} {\bibfield  {journal} {\bibinfo  {journal} {Phys. Rev. A}\ }\textbf {\bibinfo {volume} {100}},\ \bibinfo {pages} {053604} (\bibinfo {year} {2019})}\BibitemShut {NoStop}%
\bibitem [{\citenamefont {Madroñero}\ \emph {et~al.}(2024)\citenamefont {Madroñero}, \citenamefont {Castro},\ and\ \citenamefont {Paredes}}]{madronero2024}%
  \BibitemOpen
  \bibfield  {author} {\bibinfo {author} {\bibfnamefont {C.}~\bibnamefont {Madroñero}}, \bibinfo {author} {\bibfnamefont {G.~A.~D.}\ \bibnamefont {Castro}},\ and\ \bibinfo {author} {\bibfnamefont {R.}~\bibnamefont {Paredes}},\ }\href@noop {} {\  (\bibinfo {year} {2024})},\ \Eprint {https://arxiv.org/abs/2405.00811} {arXiv:2405.00811} \BibitemShut {NoStop}%
\bibitem [{\citenamefont {Zeng}\ \emph {et~al.}(2024)\citenamefont {Zeng}, \citenamefont {Zhu},\ and\ \citenamefont {He}}]{zeng2024}%
  \BibitemOpen
  \bibfield  {author} {\bibinfo {author} {\bibfnamefont {J.-H.}\ \bibnamefont {Zeng}}, \bibinfo {author} {\bibfnamefont {Q.}~\bibnamefont {Zhu}},\ and\ \bibinfo {author} {\bibfnamefont {L.}~\bibnamefont {He}},\ }\href@noop {} {\  (\bibinfo {year} {2024})},\ \Eprint {https://arxiv.org/abs/2405.20732} {arXiv:2405.20732} \BibitemShut {NoStop}%
\bibitem [{\citenamefont {Wang}\ \emph {et~al.}(2024)\citenamefont {Wang}, \citenamefont {Gao}, \citenamefont {Zhang}, \citenamefont {Zhai},\ and\ \citenamefont {Shi}}]{GaoChao2024}%
  \BibitemOpen
  \bibfield  {author} {\bibinfo {author} {\bibfnamefont {C.}~\bibnamefont {Wang}}, \bibinfo {author} {\bibfnamefont {C.}~\bibnamefont {Gao}}, \bibinfo {author} {\bibfnamefont {J.}~\bibnamefont {Zhang}}, \bibinfo {author} {\bibfnamefont {H.}~\bibnamefont {Zhai}},\ and\ \bibinfo {author} {\bibfnamefont {Z.-Y.}\ \bibnamefont {Shi}},\ }\href@noop {} {\bibfield  {journal} {\bibinfo  {journal} {Phys. Rev. Lett.}\ }\textbf {\bibinfo {volume} {133}},\ \bibinfo {pages} {163401} (\bibinfo {year} {2024})}\BibitemShut {NoStop}%
\bibitem [{\citenamefont {Tian}\ \emph {et~al.}(2024)\citenamefont {Tian}, \citenamefont {Zhang}, \citenamefont {Wu}, \citenamefont {Liu}, \citenamefont {Zhang}, \citenamefont {Li},\ and\ \citenamefont {Liu}}]{tian2024}%
  \BibitemOpen
  \bibfield  {author} {\bibinfo {author} {\bibfnamefont {R.}~\bibnamefont {Tian}}, \bibinfo {author} {\bibfnamefont {Y.}~\bibnamefont {Zhang}}, \bibinfo {author} {\bibfnamefont {T.}~\bibnamefont {Wu}}, \bibinfo {author} {\bibfnamefont {M.}~\bibnamefont {Liu}}, \bibinfo {author} {\bibfnamefont {Y.-C.}\ \bibnamefont {Zhang}}, \bibinfo {author} {\bibfnamefont {S.}~\bibnamefont {Li}},\ and\ \bibinfo {author} {\bibfnamefont {B.}~\bibnamefont {Liu}},\ }\href@noop {} {\  (\bibinfo {year} {2024})},\ \Eprint {https://arxiv.org/abs/2407.21466} {arXiv:2407.21466} \BibitemShut {NoStop}%
\bibitem [{\citenamefont {Ho}(1998)}]{ho1998}%
  \BibitemOpen
  \bibfield  {author} {\bibinfo {author} {\bibfnamefont {T.-L.}\ \bibnamefont {Ho}},\ }\href@noop {} {\bibfield  {journal} {\bibinfo  {journal} {Phys. Rev. Lett.}\ }\textbf {\bibinfo {volume} {81}},\ \bibinfo {pages} {742} (\bibinfo {year} {1998})}\BibitemShut {NoStop}%
\bibitem [{\citenamefont {Ohmi}\ and\ \citenamefont {Machida}(1998)}]{ohmi1998}%
  \BibitemOpen
  \bibfield  {author} {\bibinfo {author} {\bibfnamefont {T.}~\bibnamefont {Ohmi}}\ and\ \bibinfo {author} {\bibfnamefont {K.}~\bibnamefont {Machida}},\ }\href@noop {} {\bibfield  {journal} {\bibinfo  {journal} {J. Phys. Soc. Jpn.}\ }\textbf {\bibinfo {volume} {67}},\ \bibinfo {pages} {1822} (\bibinfo {year} {1998})}\BibitemShut {NoStop}%
\bibitem [{\citenamefont {Stenger}\ \emph {et~al.}(1998)\citenamefont {Stenger}, \citenamefont {Inouye}, \citenamefont {{Stamper-Kurn}}, \citenamefont {Miesner}, \citenamefont {Chikkatur},\ and\ \citenamefont {Ketterle}}]{stenger1998}%
  \BibitemOpen
  \bibfield  {author} {\bibinfo {author} {\bibfnamefont {J.}~\bibnamefont {Stenger}}, \bibinfo {author} {\bibfnamefont {S.}~\bibnamefont {Inouye}}, \bibinfo {author} {\bibfnamefont {D.~M.}\ \bibnamefont {{Stamper-Kurn}}}, \bibinfo {author} {\bibfnamefont {H.-J.}\ \bibnamefont {Miesner}}, \bibinfo {author} {\bibfnamefont {A.~P.}\ \bibnamefont {Chikkatur}},\ and\ \bibinfo {author} {\bibfnamefont {W.}~\bibnamefont {Ketterle}},\ }\href@noop {} {\bibfield  {journal} {\bibinfo  {journal} {Nature}\ }\textbf {\bibinfo {volume} {396}},\ \bibinfo {pages} {345} (\bibinfo {year} {1998})}\BibitemShut {NoStop}%
\bibitem [{\citenamefont {Park}\ \emph {et~al.}(2021)\citenamefont {Park}, \citenamefont {Cao}, \citenamefont {Watanabe}, \citenamefont {Taniguchi},\ and\ \citenamefont {Jarillo-Herrero}}]{park2021}%
  \BibitemOpen
  \bibfield  {author} {\bibinfo {author} {\bibfnamefont {J.~M.}\ \bibnamefont {Park}}, \bibinfo {author} {\bibfnamefont {Y.}~\bibnamefont {Cao}}, \bibinfo {author} {\bibfnamefont {K.}~\bibnamefont {Watanabe}}, \bibinfo {author} {\bibfnamefont {T.}~\bibnamefont {Taniguchi}},\ and\ \bibinfo {author} {\bibfnamefont {P.}~\bibnamefont {Jarillo-Herrero}},\ }\href@noop {} {\bibfield  {journal} {\bibinfo  {journal} {Nature}\ }\textbf {\bibinfo {volume} {590}},\ \bibinfo {pages} {249} (\bibinfo {year} {2021})}\BibitemShut {NoStop}%
\bibitem [{\citenamefont {Zhu}\ \emph {et~al.}(2020)\citenamefont {Zhu}, \citenamefont {Carr}, \citenamefont {Massatt}, \citenamefont {Luskin},\ and\ \citenamefont {Kaxiras}}]{zhu2020}%
  \BibitemOpen
  \bibfield  {author} {\bibinfo {author} {\bibfnamefont {Z.}~\bibnamefont {Zhu}}, \bibinfo {author} {\bibfnamefont {S.}~\bibnamefont {Carr}}, \bibinfo {author} {\bibfnamefont {D.}~\bibnamefont {Massatt}}, \bibinfo {author} {\bibfnamefont {M.}~\bibnamefont {Luskin}},\ and\ \bibinfo {author} {\bibfnamefont {E.}~\bibnamefont {Kaxiras}},\ }\href@noop {} {\bibfield  {journal} {\bibinfo  {journal} {Phys. Rev. Lett.}\ }\textbf {\bibinfo {volume} {125}},\ \bibinfo {pages} {116404} (\bibinfo {year} {2020})}\BibitemShut {NoStop}%
\bibitem [{\citenamefont {Bai}\ \emph {et~al.}(2023)\citenamefont {Bai}, \citenamefont {Li}, \citenamefont {Liu}, \citenamefont {Guo}, \citenamefont {Pack}, \citenamefont {Wang}, \citenamefont {Dean}, \citenamefont {Hone},\ and\ \citenamefont {Zhu}}]{bai2023}%
  \BibitemOpen
  \bibfield  {author} {\bibinfo {author} {\bibfnamefont {Y.}~\bibnamefont {Bai}}, \bibinfo {author} {\bibfnamefont {Y.}~\bibnamefont {Li}}, \bibinfo {author} {\bibfnamefont {S.}~\bibnamefont {Liu}}, \bibinfo {author} {\bibfnamefont {Y.}~\bibnamefont {Guo}}, \bibinfo {author} {\bibfnamefont {J.}~\bibnamefont {Pack}}, \bibinfo {author} {\bibfnamefont {J.}~\bibnamefont {Wang}}, \bibinfo {author} {\bibfnamefont {C.~R.}\ \bibnamefont {Dean}}, \bibinfo {author} {\bibfnamefont {J.}~\bibnamefont {Hone}},\ and\ \bibinfo {author} {\bibfnamefont {X.}~\bibnamefont {Zhu}},\ }\href@noop {} {\bibfield  {journal} {\bibinfo  {journal} {Nano Lett.}\ }\textbf {\bibinfo {volume} {23}},\ \bibinfo {pages} {11621} (\bibinfo {year} {2023})}\BibitemShut {NoStop}%
\bibitem [{\citenamefont {He}\ \emph {et~al.}(2024)\citenamefont {He}, \citenamefont {Gong}, \citenamefont {Tong}, \citenamefont {Zhai}, \citenamefont {Yao},\ and\ \citenamefont {An}}]{yao2024topological}%
  \BibitemOpen
  \bibfield  {author} {\bibinfo {author} {\bibfnamefont {H.}~\bibnamefont {He}}, \bibinfo {author} {\bibfnamefont {Z.}~\bibnamefont {Gong}}, \bibinfo {author} {\bibfnamefont {Q.-J.}\ \bibnamefont {Tong}}, \bibinfo {author} {\bibfnamefont {D.}~\bibnamefont {Zhai}}, \bibinfo {author} {\bibfnamefont {W.}~\bibnamefont {Yao}},\ and\ \bibinfo {author} {\bibfnamefont {X.-T.}\ \bibnamefont {An}},\ }\href@noop {} {\  (\bibinfo {year} {2024})},\ \Eprint {https://arxiv.org/abs/2410.05197} {arXiv:2410.05197} \BibitemShut {NoStop}%
\bibitem [{\citenamefont {Zurek}(1985)}]{zurek1985}%
  \BibitemOpen
  \bibfield  {author} {\bibinfo {author} {\bibfnamefont {W.~H.}\ \bibnamefont {Zurek}},\ }\href@noop {} {\bibfield  {journal} {\bibinfo  {journal} {Nature}\ }\textbf {\bibinfo {volume} {317}},\ \bibinfo {pages} {505} (\bibinfo {year} {1985})}\BibitemShut {NoStop}%
\bibitem [{\citenamefont {Damski}(2005)}]{Damski2005}%
  \BibitemOpen
  \bibfield  {author} {\bibinfo {author} {\bibfnamefont {B.}~\bibnamefont {Damski}},\ }\href {https://doi.org/10.1103/PhysRevLett.95.035701} {\bibfield  {journal} {\bibinfo  {journal} {Phys. Rev. Lett.}\ }\textbf {\bibinfo {volume} {95}},\ \bibinfo {pages} {035701} (\bibinfo {year} {2005})}\BibitemShut {NoStop}%
\bibitem [{\citenamefont {Damski}\ and\ \citenamefont {Zurek}(2007)}]{Damski2007}%
  \BibitemOpen
  \bibfield  {author} {\bibinfo {author} {\bibfnamefont {B.}~\bibnamefont {Damski}}\ and\ \bibinfo {author} {\bibfnamefont {W.~H.}\ \bibnamefont {Zurek}},\ }\href@noop {} {\bibfield  {journal} {\bibinfo  {journal} {Phys. Rev. Lett.}\ }\textbf {\bibinfo {volume} {99}},\ \bibinfo {pages} {130402} (\bibinfo {year} {2007})}\BibitemShut {NoStop}%
\bibitem [{\citenamefont {Anquez}\ \emph {et~al.}(2016)\citenamefont {Anquez}, \citenamefont {Robbins}, \citenamefont {Bharath}, \citenamefont {Boguslawski}, \citenamefont {Hoang},\ and\ \citenamefont {Chapman}}]{Anquez2016}%
  \BibitemOpen
  \bibfield  {author} {\bibinfo {author} {\bibfnamefont {M.}~\bibnamefont {Anquez}}, \bibinfo {author} {\bibfnamefont {B.~A.}\ \bibnamefont {Robbins}}, \bibinfo {author} {\bibfnamefont {H.~M.}\ \bibnamefont {Bharath}}, \bibinfo {author} {\bibfnamefont {M.}~\bibnamefont {Boguslawski}}, \bibinfo {author} {\bibfnamefont {T.~M.}\ \bibnamefont {Hoang}},\ and\ \bibinfo {author} {\bibfnamefont {M.~S.}\ \bibnamefont {Chapman}},\ }\href@noop {} {\bibfield  {journal} {\bibinfo  {journal} {Phys. Rev. Lett.}\ }\textbf {\bibinfo {volume} {116}},\ \bibinfo {pages} {155301} (\bibinfo {year} {2016})}\BibitemShut {NoStop}%
\bibitem [{\citenamefont {Petrov}\ \emph {et~al.}(2000)\citenamefont {Petrov}, \citenamefont {Holzmann},\ and\ \citenamefont {Shlyapnikov}}]{2Dtrap2000}%
  \BibitemOpen
  \bibfield  {author} {\bibinfo {author} {\bibfnamefont {D.~S.}\ \bibnamefont {Petrov}}, \bibinfo {author} {\bibfnamefont {M.}~\bibnamefont {Holzmann}},\ and\ \bibinfo {author} {\bibfnamefont {G.~V.}\ \bibnamefont {Shlyapnikov}},\ }\href@noop {} {\bibfield  {journal} {\bibinfo  {journal} {Phys. Rev. Lett.}\ }\textbf {\bibinfo {volume} {84}},\ \bibinfo {pages} {2551} (\bibinfo {year} {2000})}\BibitemShut {NoStop}%
\bibitem [{\citenamefont {Kawaguchi}\ and\ \citenamefont {Ueda}(2012)}]{kawaguchi2012}%
  \BibitemOpen
  \bibfield  {author} {\bibinfo {author} {\bibfnamefont {Y.}~\bibnamefont {Kawaguchi}}\ and\ \bibinfo {author} {\bibfnamefont {M.}~\bibnamefont {Ueda}},\ }\href@noop {} {\bibfield  {journal} {\bibinfo  {journal} {Phys. Rep.}\ }\textbf {\bibinfo {volume} {520}},\ \bibinfo {pages} {253} (\bibinfo {year} {2012})}\BibitemShut {NoStop}%
\bibitem [{\citenamefont {Tong}\ \emph {et~al.}(2017)\citenamefont {Tong}, \citenamefont {Yu}, \citenamefont {Zhu}, \citenamefont {Wang}, \citenamefont {Xu},\ and\ \citenamefont {Yao}}]{tong2017}%
  \BibitemOpen
  \bibfield  {author} {\bibinfo {author} {\bibfnamefont {Q.}~\bibnamefont {Tong}}, \bibinfo {author} {\bibfnamefont {H.}~\bibnamefont {Yu}}, \bibinfo {author} {\bibfnamefont {Q.}~\bibnamefont {Zhu}}, \bibinfo {author} {\bibfnamefont {Y.}~\bibnamefont {Wang}}, \bibinfo {author} {\bibfnamefont {X.}~\bibnamefont {Xu}},\ and\ \bibinfo {author} {\bibfnamefont {W.}~\bibnamefont {Yao}},\ }\href {https://doi.org/10.1038/nphys3968} {\bibfield  {journal} {\bibinfo  {journal} {Nat. Phys.}\ }\textbf {\bibinfo {volume} {13}},\ \bibinfo {pages} {356} (\bibinfo {year} {2017})}\BibitemShut {NoStop}%
\bibitem [{\citenamefont {Sadler}\ \emph {et~al.}(2006)\citenamefont {Sadler}, \citenamefont {Higbie}, \citenamefont {Leslie}, \citenamefont {Vengalattore},\ and\ \citenamefont {{Stamper-Kurn}}}]{sadler2006}%
  \BibitemOpen
  \bibfield  {author} {\bibinfo {author} {\bibfnamefont {L.~E.}\ \bibnamefont {Sadler}}, \bibinfo {author} {\bibfnamefont {J.~M.}\ \bibnamefont {Higbie}}, \bibinfo {author} {\bibfnamefont {S.~R.}\ \bibnamefont {Leslie}}, \bibinfo {author} {\bibfnamefont {M.}~\bibnamefont {Vengalattore}},\ and\ \bibinfo {author} {\bibfnamefont {D.~M.}\ \bibnamefont {{Stamper-Kurn}}},\ }\href@noop {} {\bibfield  {journal} {\bibinfo  {journal} {Nature}\ }\textbf {\bibinfo {volume} {443}},\ \bibinfo {pages} {312} (\bibinfo {year} {2006})}\BibitemShut {NoStop}%
\bibitem [{\citenamefont {Saito}\ \emph {et~al.}(2007)\citenamefont {Saito}, \citenamefont {Kawaguchi},\ and\ \citenamefont {Ueda}}]{saito2007}%
  \BibitemOpen
  \bibfield  {author} {\bibinfo {author} {\bibfnamefont {H.}~\bibnamefont {Saito}}, \bibinfo {author} {\bibfnamefont {Y.}~\bibnamefont {Kawaguchi}},\ and\ \bibinfo {author} {\bibfnamefont {M.}~\bibnamefont {Ueda}},\ }\href {https://doi.org/10.1103/PhysRevA.76.043613} {\bibfield  {journal} {\bibinfo  {journal} {Phys. Rev. A}\ }\textbf {\bibinfo {volume} {76}},\ \bibinfo {pages} {043613} (\bibinfo {year} {2007})}\BibitemShut {NoStop}%
\bibitem [{\citenamefont {Widera}\ \emph {et~al.}(2005)\citenamefont {Widera}, \citenamefont {Gerbier}, \citenamefont {F\"olling}, \citenamefont {Gericke}, \citenamefont {Mandel},\ and\ \citenamefont {Bloch}}]{widera2005}%
  \BibitemOpen
  \bibfield  {author} {\bibinfo {author} {\bibfnamefont {A.}~\bibnamefont {Widera}}, \bibinfo {author} {\bibfnamefont {F.}~\bibnamefont {Gerbier}}, \bibinfo {author} {\bibfnamefont {S.}~\bibnamefont {F\"olling}}, \bibinfo {author} {\bibfnamefont {T.}~\bibnamefont {Gericke}}, \bibinfo {author} {\bibfnamefont {O.}~\bibnamefont {Mandel}},\ and\ \bibinfo {author} {\bibfnamefont {I.}~\bibnamefont {Bloch}},\ }\href {https://doi.org/10.1103/PhysRevLett.95.190405} {\bibfield  {journal} {\bibinfo  {journal} {Phys. Rev. Lett.}\ }\textbf {\bibinfo {volume} {95}},\ \bibinfo {pages} {190405} (\bibinfo {year} {2005})}\BibitemShut {NoStop}%
\bibitem [{\citenamefont {Zhang}\ \emph {et~al.}(2005)\citenamefont {Zhang}, \citenamefont {Zhou}, \citenamefont {Chang}, \citenamefont {Chapman},\ and\ \citenamefont {You}}]{zhang2005}%
  \BibitemOpen
  \bibfield  {author} {\bibinfo {author} {\bibfnamefont {W.}~\bibnamefont {Zhang}}, \bibinfo {author} {\bibfnamefont {D.~L.}\ \bibnamefont {Zhou}}, \bibinfo {author} {\bibfnamefont {M.-S.}\ \bibnamefont {Chang}}, \bibinfo {author} {\bibfnamefont {M.~S.}\ \bibnamefont {Chapman}},\ and\ \bibinfo {author} {\bibfnamefont {L.}~\bibnamefont {You}},\ }\href {https://doi.org/10.1103/PhysRevA.72.013602} {\bibfield  {journal} {\bibinfo  {journal} {Phys. Rev. A}\ }\textbf {\bibinfo {volume} {72}},\ \bibinfo {pages} {013602} (\bibinfo {year} {2005})}\BibitemShut {NoStop}%
\bibitem [{\citenamefont {Mahmud}\ and\ \citenamefont {Tiesinga}(2013)}]{mahumd2013}%
  \BibitemOpen
  \bibfield  {author} {\bibinfo {author} {\bibfnamefont {K.~W.}\ \bibnamefont {Mahmud}}\ and\ \bibinfo {author} {\bibfnamefont {E.}~\bibnamefont {Tiesinga}},\ }\href {https://doi.org/10.1103/PhysRevA.88.023602} {\bibfield  {journal} {\bibinfo  {journal} {Phys. Rev. A}\ }\textbf {\bibinfo {volume} {88}},\ \bibinfo {pages} {023602} (\bibinfo {year} {2013})}\BibitemShut {NoStop}%
\bibitem [{\citenamefont {Stamper-Kurn}\ and\ \citenamefont {Ueda}(2013)}]{dan2013}%
  \BibitemOpen
  \bibfield  {author} {\bibinfo {author} {\bibfnamefont {D.~M.}\ \bibnamefont {Stamper-Kurn}}\ and\ \bibinfo {author} {\bibfnamefont {M.}~\bibnamefont {Ueda}},\ }\href {https://doi.org/10.1103/RevModPhys.85.1191} {\bibfield  {journal} {\bibinfo  {journal} {Rev. Mod. Phys.}\ }\textbf {\bibinfo {volume} {85}},\ \bibinfo {pages} {1191} (\bibinfo {year} {2013})}\BibitemShut {NoStop}%
\bibitem [{\citenamefont {Zhu}\ \emph {et~al.}(2015)\citenamefont {Zhu}, \citenamefont {Sun},\ and\ \citenamefont {Wu}}]{zhu2015}%
  \BibitemOpen
  \bibfield  {author} {\bibinfo {author} {\bibfnamefont {Q.}~\bibnamefont {Zhu}}, \bibinfo {author} {\bibfnamefont {Q.-f.}\ \bibnamefont {Sun}},\ and\ \bibinfo {author} {\bibfnamefont {B.}~\bibnamefont {Wu}},\ }\href@noop {} {\bibfield  {journal} {\bibinfo  {journal} {Phys. Rev. A}\ }\textbf {\bibinfo {volume} {91}},\ \bibinfo {pages} {023633} (\bibinfo {year} {2015})}\BibitemShut {NoStop}%
\end{thebibliography}%
\end{document}